\documentclass{elsart}
\usepackage{natbib}
\usepackage{graphicx}
\usepackage{amssymb}
\usepackage{amsmath}
\usepackage{bm}

\def\url#1{{\ttfamily\def\/{/\discretionary{}{}{}}#1}}
\def\bibcode#1{}

\begin{document}
\begin{frontmatter}
\title{Simulations of Baryon Oscillations}
\author{Eric Huff${}^1$, A.E. Schulz${}^1$, Martin White${}^{1,2}$}
\author{David J. Schlegel${}^2$, Michael S. Warren${}^3$}
\address{${}^1$Departments of Physics and Astronomy,\\
University of California, Berkeley, CA 94720}
\address{${}^2$Lawrence Berkeley National Laboratory,\\
1 Cyclotron Road, Berkeley, CA}
\address{${}^3$Theoretical Astrophysics Division, Los Alamos National
Laboratory}
\thanks[ehemail]{ehuff@astro.berkeley.edu}
\thanks[asemail]{aschulz@astro.berkeley.edu}
\thanks[mwemail]{mwhite@berkeley.edu}
\thanks[dsemail]{djschlegel@lbl.gov}
\thanks[wwemail]{msw@lanl.gov}

\begin{abstract}
The coupling of photons and baryons by Thomson scattering in the early
universe imprints features in both the Cosmic Microwave Background (CMB)
and matter power spectra.  The former have been used to constrain a host
of cosmological parameters, the latter have the potential to strongly
constrain the expansion history of the universe and dark energy.
Key to this program is the means to localize the primordial features in
observations of galaxy spectra which necessarily involve galaxy bias,
non-linear evolution and redshift space distortions.  We present calculations,
based on mock catalogs produced from high-resolution N-body simulations,
which show the range of behaviors we might expect of galaxies in the real
universe.  We investigate physically motivated fitting forms which include
the effects of non-linearity, galaxy bias and redshift space distortions and
discuss methods for analysis of upcoming data.
In agreement with earlier work, we find that a survey of several Gpc${}^3$
would constrain the sound horizon at $z\sim 1$ to about 1\%.
\end{abstract}
\end{frontmatter}

\section{Introduction}

Recently four groups \cite{SDSS}, using data from the
Sloan Digital Sky Survey\footnote{http://www.sdss.org/}, published evidence
for features in the matter power spectrum on scales of $100\,$Mpc.
These features, long predicted, hold the promise of another route to
understanding the expansion history of the universe and the influence
of dark energy \cite{EisReview}.

Oscillations in the baryon-photon fluid at $z\sim 10^3$ lead to a
series of almost harmonic peaks in the matter power spectrum, or a
bump in the correlation function, arising from a preferred scale in
the universe: the sound horizon.
(A description of the physics leading to the features can be found in
\cite{EHSS} or Appendix A of \cite{MeiWhiPea}; a comparison of the Fourier
and configuration space pictures is presented in \cite{ESW06}.)
It was pointed out in Refs.~\cite{CooHuHutJof,Eis03} that this scale could
be used as a standard ruler to constrain the distance-redshift relation, the
expansion of the universe and dark energy.  Numerous authors
\cite{Fisher} have now observed that a high-$z$ galaxy survey\footnote{It
is even possible that such oscillations could be seen in the Ly-$\alpha$
forest \cite{Davis} or in very large cluster surveys \cite{Ang}.} covering
upwards of several hundred square degrees could place interesting
constraints on dark energy.
Key to realizing this is the ability to accurately predict the physical scale
at which the oscillations appear in the power spectrum plus the means to
localize those primordial features in observations of galaxy spectra which
necessarily involve galaxy bias, non-linear evolution and redshift space
distortions.
The former problem seems well in hand \cite{SSWZ,EisWhi}.
Preliminary investigations of the latter problem were presented in
Refs.~\cite{Ang,Millenium,PMBaryon,SeoEis05,GuzBer}.
We continue these investigations in this paper using a large set of
high resolution N-body simulations.

The outline of the paper is as follows: \S\ref{sec:sim} describes
our N-body simulations and the construction of the mock galaxy catalogs
using halo model methods.  It also presents some basic properties of the
galaxy clustering.  \S\ref{sec:models} introduces the models for
the 2-point function that we consider in this paper.  \S\ref{sec:dxi}
introduces a new configuration space band-power statistic which we believe
is useful for BAO work while \S\ref{sec:methodology} introduces our fitting
methodology.  Our primary results are described in \S\ref{sec:results}.
Some preliminary investigations of the reconstruction method of \cite{ESSS06}
are described in \S\ref{sec:reconstruction} and our conclusions are
presented in \S\ref{sec:conclusions}.

\section{Simulations} \label{sec:sim}

We need an ``event generator'' which can be used to develop methods for
going from observations of galaxies to cosmology.  Ideally this tool
would encode many of the complications we expect in real observations
but be based on a known cosmology.  To this end we use a series of
simulations of a $\Lambda$CDM cosmology ($\Omega_M=0.3=1-\Omega_\Lambda$,
$\Omega_B=0.046$, $h=0.7$, $n=1$ and $\sigma_8=0.9$).  The linear theory
mass spectrum was computed by evolution of the coupled Einstein, fluid
and Boltzmann equations using the code described in \cite{Boltz}.  (A
comparison of this code to CMBfast \cite{CMBfast} is given in \cite{SSWZ}.)
For this model the sound horizon\footnote{We caution the reader that the
definition of the sound horizon, like that of the epoch of last scattering,
is one of convention.  Unfortunately several conventions exist in the
literature.  Along with fitting formulae of limited accuracy this makes it
difficult to compare quoted numbers at the percent level.  We define $s$ as
the integral of the sound speed up to the redshift where $\tau=1$, excluding
the contribution from $z<z_{\rm reion}$.}, $s=143\,$Mpc or $100.2\,h^{-1}$Mpc.

For tests in which we need large numbers of runs (i.e.~computing covariance
matrices) we use mock catalogs based on Gaussian density fields.
We first generate a Gaussian density field with the desired power spectrum
(in our case the linear theory spectrum) on a $512^3$ grid in a box of side
$1.1\,h^{-1}$Gpc.  A `galaxy' is then placed at random in the cell with a
probability proportional to $\exp[3\nu-0.3\nu^2]$ for $\nu\ge 0$ and
$\exp[3\nu]$ for $\nu<0$, where $\nu=\delta/\sigma$ is the peak height.
Similar techniques are used to construct mock galaxy redshift surveys in
\cite{CHWF}.  The non-linear mapping of the Gaussian density field mocks up
the action of gravity, inducing extra power on small scales and correlating
different scales in Fourier space.  The resulting `galaxy' field has a power
spectrum with roughly constant bias on large scales and excess power on small
scales, though the form does not match in detail the more realistic catalogs
produced with the halos found in N-body simulations.

A more realistic catalog can be produced using N-body simulations which have
enough spatial and force resolution to resolve the halos hosting the galaxies
of interest for BAO surveys.
The basis for these calculations is a sequence of high force resolution
N-body simulations in a periodic, cubical box of side $1.1\,h^{-1}$Gpc.
These simulations were carried out with the HOT \cite{HOT} and TreePM
\cite{TreePM} codes.
In all we ran 3 simulations, with different randomly generated Gaussian
initial conditions, which evolved $1024^3$ particles of mass
$10^{11}\,h^{-1}M_\odot$ from $z=34$ to $z=1$.
The Plummer softening was $35\,h^{-1}$kpc (comoving).

\begin{figure}
\begin{center}
\resizebox{5.5in}{!}{\includegraphics{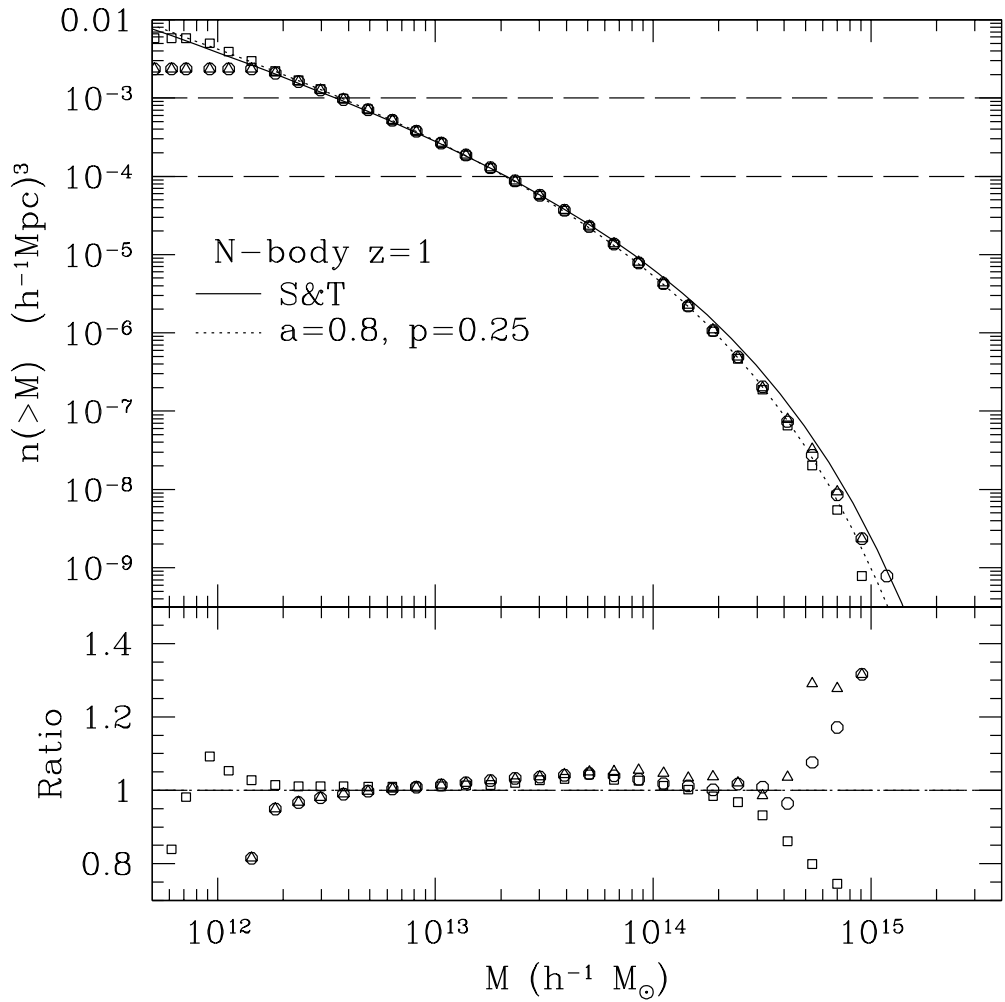}}
\end{center}
\caption{(Upper) The cumulative mass function of halos from the 3 simulations
at $z=1$.  We define $M$ as $1.03\times$ the sum of the masses of the particles
in the FoF group.  The last point plotted in each case is the mass of the
largest halo in the box.  The solid line shows the fit of
Ref.~\protect\cite{ST} while the dotted line shows the results with $a=0.8$
and $p=0.25$ as discussed in the text.  The horizontal dashed lines show the
range of number densities of most interest for this work.  (Lower) The ratio
of the N-body results to the fit with $a=0.8$ and $p=0.25$.}
\label{fig:massfn}
\end{figure}

The simulations were chosen to be the largest we could practically run
several realizations of, while retaining sufficient force resolution to
resolve the halos likely to host the galaxies of interest.  This allowed
us to simulate a volume comparable to that of proposed future surveys
\cite{WFMOS} at $z=1$.
While observational campaigns could also study baryon oscillations at
higher redshift (e.g.~$z=3$) going to earlier times becomes increasingly
difficult for simulations.  The volume for a given survey area increases and
the characteristic mass scale of halos decreases to earlier times, making the
required dynamic range infeasible for direct simulation at present.  Thus
we will focus our attention here on $z=1$.

For each output we generate a catalog of halos using the Friends-of-Friends
algorithm \cite{FoF} with a linking length of $b=0.175$ in units of the
mean inter-particle spacing.
This procedure partitions the particles into equivalence classes, by linking
together all particle pairs separated by less than a distance $b$.
The groups correspond roughly to all particles above a density of about
$3/(2\pi b^3)\simeq 90$ times the background density and we keep all
groups with more than 16 particles.  Increasing the ``friends-of-friends''
mass of the groups by a few percent gives a good match to the
analytic mass function of Ref.~\cite{ST}.  However we find that even with
this increase the $z=1$ results are better fit if we change the parameters
in the analytic mass function ($p$ which controls the low mass slope of the
mass function and $a$ which controls the exponential suppression at high mass)
from $a=0.707$ and $p=0.3$ to $a=0.8$ and $p=0.25$.
Over the range $2\times 10^{12}\le M\le 2\times 10^{14}\,h^{-1}M_\odot$ the
resulting fit is good to $5\%$ in number density (see Fig.~\ref{fig:massfn}).

\begin{table}
\begin{center}
\begin{tabular}{cccc|cccc}
\multicolumn{4}{c|}{$\bar{n}=10^{-3.0}\,h^3\,{\rm Mpc}^{-3}$} &
\multicolumn{4}{c}{$\bar{n}=10^{-3.5}\,h^3\,{\rm Mpc}^{-3}$} \\
$\log_{10}M_{\rm min}$ & $\log_{10}M_1$ &
  $\langle b\rangle$   & $\log_{10}\langle M\rangle$ &
$\log_{10}M_{\rm min}$ & $\log_{10}M_1$ &
  $\langle b\rangle$   & $\log_{10}\langle M\rangle$ \\
  12.8                 & 13.0           & 2.4 & 13.6 &
  13.3                 & 13.0           & 3.1 & 13.9 \\
  12.7                 & 13.5           & 1.9 & 13.3 &
  13.1                 & 13.5           & 2.5 & 13.7 \\
  12.6                 & 14.0           & 1.8 & 13.1 &
  13.0                 & 14.0           & 2.3 & 13.5 \\
  12.6                 & 14.5           & 1.7 & 13.1 &
  13.0                 & 14.5           & 2.1 & 13.4
\end{tabular}
\end{center}
\caption{The HOD parameters of Eq.~\protect\ref{eqn:hod} for some of our
catalogs at $z=1$.  Masses are in units of $h^{-1}M_\odot$.  Within each
set the catalogs have a fixed space density, $\bar{n}$.  The large-scale
bias, $\langle b\rangle$, and the galaxy weighted mean halo mass,
$\langle M\rangle$, are also listed.}
\label{tab:hod10}
\end{table}

We make mock catalogs using two different procedures.  First we apply
a simple threshold mass to the group catalogs, taking our tracers to be
the central galaxies of halos above the mass threshold.
To allow the inclusion of satellites we choose a mean occupancy of halos:
$N(M)\equiv\left\langle N_{\rm gal}(M_{\rm halo})\right\rangle$.
Each halo either hosts a central galaxy or does not.
For each halo we define a galaxy to live at the minimum of the halo
potential with probability $p={\rm min}[1,N(M)]$.
The central galaxy inherits the velocity of the halo, which we take to
be the average velocity of the halo particles weighted by the square of the
potential.  This weighting emphasizes particles near the halo center and
allows the central galaxy to move with respect to the center-of-mass (com)
of the halo, but the difference between the com velocity and the central
galaxy velocity is typically only a few tens of km$\,{\rm s}^{-1}$.
Following Ref.~\cite{KBWKGAP}, if $N(M)>1$ the mean number of satellites,
$N_{\rm sat}=N(M)-1$, is computed for the halo and a Poisson random number,
$n_{\rm sat}$, drawn.
Then $n_{\rm sat}$ dark matter particles, chosen at random, are anointed as
galaxies.  Our fiducial model thus has the satellite galaxies tracing the
dark matter in the halo.
The galaxy velocity is taken to be the particle velocity, thus the satellites
have no velocity bias.  However since the central galaxy is nearly at rest
with respect to the halo, the population as a whole has a different velocity
field than the dark matter.
The characteristic mass, $M_\star$, for our models at $z=1$ is
$2\times 10^{11}\,h^{-1}M_\odot$.  As all of our tracer galaxies live in
halos more massive than $M_\star$ they have biases greater than 1.

We follow \cite{PMBaryon} and choose a simple two-parameter form for $N(M)$:
\begin{equation}
  N(M) = \Theta(M-M_{\rm min})\ \frac{\left(M-M_{\rm min}\right)+M_1}{M_1}
\label{eqn:hod}
\end{equation}
where $\Theta(x)$ is the Heaviside step function.
If we take $M_1\to\infty$ our catalog reduces to the catalog of halos more
massive than $M_{\rm min}$.  By holding $\bar{n}$ fixed
we can specify a 1-parameter sequence of models with varying $M_{\rm min}$,
large-scale bias, $\langle b\rangle$, or galaxy weighted mean halo mass
$\langle M\rangle$ (see Table \ref{tab:hod10}).
In comparing the models with different HODs it is important to remember that
we hold $\bar{n}$ fixed within each sequence, so variations in mean halo mass,
satellite fraction, bias etc are highly correlated.
To test the dependence on the slope of the satellite contribution we
also ran some models where $n_{\rm sat}\propto M^{1/2}$.  The parameters of
these models are listed in Table \ref{tab:hod05}.
Both theoretical \cite{ConWecKra} and observational \cite{YanMadWhi}
results suggest that at higher redshift $M_{\rm min}\approx M_1$.  These
models have the larger biases, mean halo masses and satellite fractions.

\begin{table}
\begin{center}
\begin{tabular}{cccc|cccc}
\multicolumn{4}{c|}{$\bar{n}=10^{-3.0}\,h^3\,{\rm Mpc}^{-3}$} &
\multicolumn{4}{c}{$\bar{n}=10^{-3.5}\,h^3\,{\rm Mpc}^{-3}$} \\
$\log_{10}M_{\rm min}$ & $\log_{10}M_1$ &
  $\langle b\rangle$   & $\log_{10}\langle M\rangle$ &
$\log_{10}M_{\rm min}$ & $\log_{10}M_1$ &
  $\langle b\rangle$   & $\log_{10}\langle M\rangle$ \\
  12.8                 & 13.0           & 2.0 & 13.3 &
  13.2                 & 13.0           & 2.6 & 13.7 \\
  12.7                 & 13.5           & 1.9 & 13.2 &
  13.1                 & 13.5           & 2.4 & 13.6 \\
  12.6                 & 14.0           & 1.8 & 13.2 &
  13.0                 & 14.0           & 2.3 & 13.5 \\
  12.6                 & 14.5           & 1.8 & 13.1 &
  13.0                 & 14.5           & 2.2 & 13.4
\end{tabular}
\end{center}
\caption{The HOD parameters used for $z=1$ catalogs with
$n_{\rm sat}\propto M^{1/2}$.  Units and labels are as
in Table \protect\ref{tab:hod10}.}
\label{tab:hod05}
\end{table}

While the galaxy models above are not prescriptive, or likely even
close to ``right'', they are physically well motivated, easy to adjust and
lead to galaxy catalogs with non-linear, scale-dependent, stochastic
biasing and redshift space distortions -- many of the complications we
will face in observations of the universe.

The statistics of counts in cubical cells allow us to infer the
stochasticity of the bias and the degree to which the 1-point
distribution of the galaxies is Poisson.  We find that the
galaxy-mass cross-correlation coefficient \cite{DekLah}, $r$, rises from
$0.6-0.7$ (depending on sample) on scales of $1\,h^{-1}$Mpc to $>95\%$ on
scales larger than $10\,h^{-1}$Mpc.  The variance of the counts divided
by their mean, which would be unity for a Poisson distribution, exceeds
one on scales $1-30\,h^{-1}$Mpc with the largest value ($\approx 20$)
on the largest scales.  This excess is easily understood as extra power
coming from large-scale clustering of the galaxies.  On Mpc scales the
value is very close to Poisson for the less biased galaxies and close to
$2$ for the more biased samples.

\section{BAO models} \label{sec:models}

The cosmological signal that interests us is primarily contained in the
two-point statistics of the galaxy density field, and we shall concentrate
on these statistics henceforth.  We begin by considering measurements in
real space, such as would be relevant to photometric or 2D surveys,
and then include redshift space distortions which are relevant for
spectroscopic surveys.  

\subsection{Real space}

There are now several models in the literature relating the (non-linear)
galaxy power spectrum to the (linear) dark matter power spectrum.
We have critically compared these to the power spectra and correlation
functions measured in our mock surveys.
To our knowledge the performance of these models in matching the shape of
the power spectra and correlation function has not been compared to mock
catalogs produced with a wide range of HOD schemes.  In particular we find
that the correlation function is a very discriminating statistic, because it
is sensitive to translinear scales.
As we discuss in \S\ref{sec:results} some of the models do not match the
clustering of our mock galaxy samples, particularly for highly biased or rare
samples.

We now describe the five models we've investigated in this paper.
The simplest is the linear bias model, which is often motivated by
arguments like those presented in \cite{SchWei}.
The linear bias model asserts that
\begin{equation}
  \Delta^2_{\rm gal}(k)=b^2 \Delta^2_{\rm lin}(\alpha k)
\label{eqn:linbias}
\end{equation} 
where $\Delta^2(k)$ denotes the dimensionless power spectrum, or variance
per $\ln k$:
\begin{equation}
  \Delta^2(k)\equiv \frac{k^3P(k)}{2\pi^2} \qquad .
\end{equation}
The parameter $b$ in Eq.~\ref{eqn:linbias} is the large scale galaxy bias
and $\Delta^2_{\rm lin}(k)$ is the linear dark matter power spectrum.
In this study we have introduced the parameter $\alpha$, which scales the
wave number in Eqs.~\ref{eqn:linbias}, \ref{eqn:lrg}, \ref{eqn:hmi}
and \ref{eqn:esw}, to parameterize small changes in the cosmology that
result in a stretch in the baryon signature.
We introduce this parameter in order to study potential degeneracies between
the other model parameters, which depend on the HOD, and the inferred
cosmology.
When $\alpha\ne 1$ the inferred length scale differs from the true length
scale, leading to an incorrect estimate of the sound horizon and hence the
constraints on dark energy.  We will use biases and errors on $\alpha$ as
an indicator of how well the sound horizon can be measured.
To translate the error in $\alpha$ into an error on dark energy parameters
we need to make further assumptions.  As a rough rule of thumb: if we assume
a constant equation of state, $w$, for the dark energy the fractional error
in $w$ is five times that in $\alpha$.
 
It has been observed several times in the literature 
\cite{PMBaryon,SeoEis05,Coo04} that the large-scale bias may not be constant
at the $1\%$ level.  Halo bias \cite{PeaksBias} will have a small scale
dependence even for weakly non-linear scales.
The distribution of galaxies within dark matter halos and halo exclusion
effects also lead to small changes in large-scale power.
The shifting of galaxy positions on $\sim 10\,h^{-1}$Mpc scales leads to
a smearing of the amplitude of the oscillations in the power spectrum.
Several attempts have been made to model these effects.

In \cite{BlaGla03} a method was introduced to empirically
fit the scale dependence, using the form 
\begin{equation}
  \Delta^2(k)= \Delta^2_{\rm ref}( k)
  \left[1+Ak\,\exp\left\{-\left(\frac{ k}{k_s}\right)^{1.4}\right\}
  \sin\left( \frac{2 \pi k}{k_A} \right) \right]
\label{eqn:blagla}
\end{equation}
Here $A$ and $k_A$ are fit parameters in the decaying sinusoid used to
characterize the baryon oscillations, $k_s\equiv 0.1\,h\,{\rm Mpc}^{-1}$
is a constant and $\Delta^2_{\rm ref}(k)$ is a reference spectrum including
the effects of Silk damping but excluding the oscillations.  The reference
spectrum is from the fitting formula in \cite{EisHu99}.
Since we are fitting non-linear power spectra we additionally allow a linear
ramp in power when doing our fits.  This approximates the broad-band power
removal suggested by \cite{BlaGla03} without correlating the errors.
The fits do prefer a positive slope to this extra factor.
In Eq.~(\ref{eqn:blagla}) a shift in $k_A$ corresponds to a shift in the
sound horizon, so $k_A$ replaces the $\alpha$ in our previous expression.
While the true spectrum cannot be accurately fit by Eq.~(\ref{eqn:blagla}),
if we concentrate around the second peak $k_A\simeq 0.058\,h{\rm Mpc}^{-1}$
provides a good fit to the peak position for our input model.
An alternative definition, used by \cite{BlaGla03}, is
$k_A=2\pi/s\simeq 0.063\,h{\rm Mpc}^{-1}$ a difference of about 10\%.
If we use the fitting function, Eq.~(26), of \cite{EisHu98} for $s$ instead
we find $k_A\simeq 0.060\,h{\rm Mpc}^{-1}$; midway between the former two
values.
The latter approximation comes closest to our best fit $k_A$ (see below) so
we shall use that -- but we note again the uncertainty in quoted values of
$s$ in the literature.
In our fitting we shall assume that $\Delta_{\rm ref}$ is known, and use the
correct cosmological parameters for our runs.

A recent analysis of the clustering of Luminous Red Galaxies (LRGs) in the
Sloan Digital Sky Survey (SDSS) instead used the fitting function
\cite{Pad06,Cole05}
\begin{equation}
  \Delta^2(k)=b^2\,\Delta_{\rm lin}^2(k) \frac{1+Qk^2}{1+ak}
  \qquad {\rm for\/} \ k<0.5\,h\,{\rm Mpc}^{-1}
\label{eqn:lrg}
\end{equation}
In this description, the parameter $Q\sim 10$ governs the scale dependence of
the bias and $a=1.7\,h^{-1}$Mpc is a constant ($a$ becomes $1.4\,h^{-1}$Mpc
in redshift space).  To test for shifts in the sound
horizon we again replace $k$ with $\alpha k$.  Note that for small $k$
this model looks like the quadratic, multiplicative bias model used in
\cite{PMBaryon}.

In \cite{ToyModel}, motivated by analytic arguments using the halo model,
the scale dependence was modeled by adding a term to the expression in
Eq.~\ref{eqn:linbias} proportional to $k^3$.  
In \cite{SeoEis05} a similar treatment was used for slightly different
reasons, involving $k^3$ times a quadratic in $k$.
For $k\ll 1$ the leading order term will dominate and the two expressions
are similar.  We consider
\begin{equation}
  \Delta_{\rm gal}^2(k) = b^2 \Delta_{\rm lin}^2(\alpha k)
   e^{-(\alpha k/k_2)^2} + \left(\alpha k/ k_1\right)^3 \qquad .
\label{eqn:hmi}
\end{equation}
The parameters $b$ and $k_1$ are the galaxy weighted large-scale halo bias
and the amplitude of the 1-halo term, and the parameter $k_2$ governs a
suppression due to halo profiles and exclusion.
We introduce $k_2$ because very little clustering power on small scales
is due to galaxies in separate halos.  It is most necessary for those
models with low $M_1$, i.e.~many satellite galaxies, the strongest 1-halo
terms, the largest mean halo mass, but the shape of the transition between
the 2- and 1-halo terms is not well constrained.  We will consider a
different transition below.
Since the baryon oscillation signal is present in $\Delta_{\rm lin}^2(k)$, it
is clear that the one halo term, which is a description of the impact of
non-linear physics, decreases the contrast of the oscillations.
Because the parameters $b$, $k_1$,and $k_2$ all depend on the HOD, the
contrast of the oscillations will depend somewhat on the type of galaxies
being selected in the BAO survey, and on the mean redshift of the survey.

Another way of viewing the effects of non-linearities and 
galaxy bias is found in \cite{ESW06}.
That analysis seeks to describe the gradual erasure of the acoustic peak
signature by considering the
distribution of Lagrangian displacements of galaxies.
The authors of \cite{ESW06} showed that to simultaneously model the smearing
due to galaxy displacement, and also the correct level of small scale power,
it is useful to add back the broad-band power that is filtered out by the
exponential suppression in Eq.~\ref{eqn:hmi}.
The fitting form from \cite{EisHu99} is again used for this purpose leading to
\begin{equation}
  \Delta_{\rm gal}^2(k) = b^2 \Delta_{\rm lin}^2(\alpha k)
  e^{-(\alpha k/k_2)^2} + \left(\alpha k/k_1\right)^3 +
  \left( 1-e^{-(\alpha k/k_2)^2} \right) b^2 \Delta^2_{\rm ref}(\alpha k)
\label{eqn:esw}
\end{equation}
The role of $k_2$ in this form differs from its role in Eq. \ref{eqn:hmi}
in that here it controls the erosion of the baryon wiggles
due to galactic motions on $\sim 10\,h^{-1}$Mpc scales, while 
preserving the overall shape of the two halo portion of the
power spectrum in the trans-linear regime.

\subsection{Redshift space}

Little work has been done on extending these models to redshift space.
In redshift space the power is enhanced, by a $z$-dependent factor, on
large scales due to supercluster infall \cite{Kaiser} and suppressed on
small scales due to virial motions within halos\footnote{This statement
assumes a $k$-space picture.  In configuration space, on large scales,
overdensities are the sites of convergent flows, so redshift space
distortions `sharpen' structure.  The correlation function thus falls
off more quickly along the line-of-sight than transverse to it.}.
To include the dependence on the angle with respect to the line of sight,
$\mu\equiv\cos\theta=\hat{k}\cdot\hat{r}$, we decompose $\Delta^2(\vec{k})$
into Legendre polynomials, $P_\ell(\mu)$, as
\begin{equation}
  \Delta^2(k,\mu) \equiv \sum_{\ell=0}^{\infty} \Delta_\ell^2(k)P_\ell(\mu)
\label{eqn:deltal}
\end{equation}
so
\begin{equation}
  \Delta_\ell^2(k) = \frac{2\ell+1}{2} \int_{-1}^{+1}d\mu\ 
  \Delta^2(k,\mu)P_\ell(\mu)
\end{equation}
Symmetry along the line-of-sight implies that the coefficients for odd $\ell$
modes vanish.

On very large scales $(k\to 0)$ supercluster infall modifies the power
spectrum as \cite{Kaiser,HamiltonReview}
\begin{equation}
  \Delta^2_{\rm red}(k,\mu) = \Delta^2_{\rm real}(k)
  \left(1+\beta\mu^2\right)^2,
\end{equation}
where $\mu\equiv\hat{k}\cdot\hat{r}$ and
$\beta\equiv f(\Omega)/b\simeq \Omega^{0.6}/b$ assuming a scale-independent
bias.  Here $f$ is the dimensionless linear growth rate of the growing mode
in linear perturbation theory which can be approximated as $\Omega^{0.6}$
\cite{Pee80}.
The coefficients of the first two multipole moments are thus
\begin{equation}
  \Delta^2_0(k) = \left(1 + \frac{2}{3}\beta +
  \frac{1}{5}\beta^2\right)\Delta_{\rm real}^2(k)
\end{equation}
and
\begin{equation}
  \Delta^2_2(k) = \left( \frac{4}{3}\beta +
  \frac{4}{7}\beta^2\right)\Delta_{\rm real}^2(k).
\end{equation}
To capture the smaller-scale angular dependence one often adds a high-$k$
cutoff; popular choices include Lorentzian, Gaussian or exponential, e.g.
\begin{equation}
  \Delta_{\rm red}^2(k,\mu)  = \Delta_{\rm real}^2(k)
  \left(1 + \beta \mu^2\right)^2 e^{-k\sigma|\mu|}
\end{equation}
We will refer to these modifications collectively as ``streaming models''
\cite{HamiltonReview}.
In general $\beta$ and $\sigma$ are regarded simply as parameters to be fit
to the data.
The analytic expressions for the resulting multipole moments are
straightforward but lengthy and will not be reproduced here.

A more empirical model for the angular dependence is explored in \cite{HatCol},
who found the quadrupole-to-monopole ratio in N-body simulations can be fit by
\begin{equation}
  \frac{\Delta^2_2}{\Delta^2_0} =
  \frac{\frac{4}{3}\beta+\frac{4}{7}\beta^2}
       {1+\frac{2}{3}\beta+\frac{1}{5}\beta^2}
  \left[1-\left(\frac{k}{k_{\rm rs}}\right)^{1.22}\right],
\label{eqn:hatcol}
\end{equation}
where $k_{\rm rs}$ is a free parameter analogous to the $\sigma$ parameter
in streaming models.  Though the authors of \cite{HatCol} do not discuss
the shape of the small-scale downturn in $\Delta_\ell^2(k)$, an additional
parameter needs to be introduced if we are to predict the run of
$\Delta_\ell^2(k)$ with $k$.  Further discussion of the interplay between
the large-scale enhancement and small-scale suppression can be found in
\cite{SheDia01,HaloRed}.

In \S\ref{sec:redresult}, we will investigate the accuracy with which each
these forms reproduces the large- and intermediate-scale
($k\le 0.2\,h\,{\rm Mpc}^{-1}$)
angular dependence of the power spectrum in redshift space.
Our approach also allows us to examine the effects of the HOD on the
redshift-space distortions, with the hope of using this additional
information to reduce degeneracies between cosmology and galaxy physics.
We do not consider in this paper how modeling of the anisotropic clustering
allows us to constrain the line-of-sight and transverse distance scales
separately.

\section{A configuration space band-power estimator} \label{sec:dxi}
 
\begin{figure}
\begin{center}
\resizebox{5.5in}{!}{\includegraphics{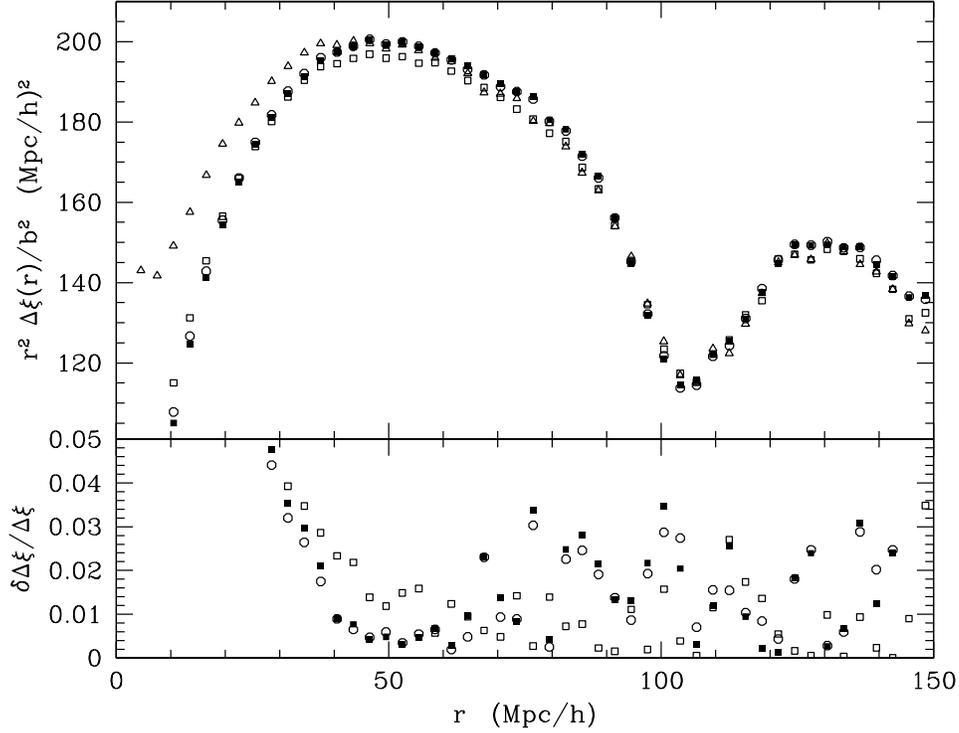}}
\end{center}
\caption{An estimate of the scale-dependence of the bias in configuration
space.  The different symbols are $\Delta\xi$ (see text) for different HOD
models (from Table~\protect\ref{tab:hod10}) each divided by a constant bias to
match near $100\,h^{-1}$Mpc.  The degree to which the shapes match indicates
how well each can be considered a constant times the dark matter correlation
function.  The lower panel shows the residuals from one of the models, taken
as a template.}
\label{fig:rbias}
\end{figure}

For each of the forms in Eqs.~\ref{eqn:linbias}-\ref{eqn:esw}
there is a corresponding model for the correlation function $\xi(r)$;
the probability, in excess of random, of finding a pair of tracers at
separation $r$.
The correlation function is related to the dimensionless power spectrum as
\begin{equation}
  \xi(r) =\int \frac{dk}{k}\ \Delta^2(k)\ j_0(kr)
  \simeq  \int \frac{dk}{k}\ \Delta^2(k)
  \left[ 1 - \frac{(kr)^2}{6} + \cdots \right]
\label{eqn:xi_wt}
\end{equation}
where $j_0$ is the zeroth spherical Bessel function and the series expansion
is valid for $kr\ll 1$.
Most of the scale dependence in the galaxy bias can be traced to the fact that
galaxies and dark matter transition from one to two halo dominance
at disparate scales \cite{ToyModel}.
In Fourier space this is spread over a range of $k$, but in configuration
space the 1-halo term is confined to scales much smaller than the scales of
interest to us.  Beyond $2-3\,h^{-1}$Mpc more than 99.9\% of the contribution
to $\xi(r)$ comes from the 2-halo term for all of our models.
For this reason we expect less scale dependence in the bias in configuration
space than in Fourier space (see also \cite{GuzBer}), as shown in
Fig.~\ref{fig:rbias}.

Formally the correlation function, $\xi(r)$, and the power spectrum, $P(k)$,
are a Fourier transform pair.
However the simulation volume is a periodic cube and our signal has support
only for $k$-modes which are integer multiples of the fundamental mode in
each dimension.
Because of this restriction the correlation function computed in the box
differs from the true continuum correlation function on scales approaching
the box size.  The modulation in power is non-trivial even on $100\,$Mpc
scales where we would like to work.
Fortunately the box is large enough to contain the modes of interest for
baryon oscillations and the difficulty is purely a technical issue.  We
choose to proceed by computing a quantity containing the same information
as $\xi(r)$ but which is less sensitive to the low-$k$ modes.
Specifically we compute
\begin{equation}
  \Delta\xi(r) \equiv \bar{\xi}(<r) - \xi(r) =
  \frac{3}{r^3}\int_0^r x^2\,dx\ \xi(x) - \xi(r)
\end{equation}
for which
\begin{equation}
  \Delta\xi(r) = \int \frac{dk}{k}\ \Delta^2(k)\ j_2(kr)
  \simeq  \int \frac{dk}{k}\ \Delta^2(k)
  \left[ \frac{(kr)^2}{15} - \frac{(kr)^4}{210} + \cdots \right]
  \label{eqn:dxi_wt}
\end{equation}
where $j_2$ is the spherical Bessel function of order $2$.
We have effectively formed a band-pass filter in $k$-space; the slowly
varying, long-wavelength contribution is significantly down-weighted while
retaining the sensitivity to $100\,$Mpc scales.
Only 3\% of the weight in the window function comes from scales below $kr=1$,
so if we include the fact that $\Delta^2(k)$ falls steeply as $k\to 0$ we see
that modes near the fundamental contribute little to $\Delta\xi(r)$.
We believe that this configuration space band-power measurement is of
considerable value in analyzing the acoustic oscillation signal.
Computing $\Delta\xi(r)$ is straightforward once $\xi(r)$ has been computed,
either in simulations or from galaxy survey volumes.  In surveys of the sky,
the correlation function is computed by assuming a mean density given by the
ratio of total object number to the survey volume.
This necessarily drives the observed correlation function to zero at survey
sized scales; a difficulty referred to as the integral constraint.
The quantity $\Delta\xi(r)$ is inherently less sensitive to the integral
problem, and is thus observationally useful apart from comparison to theory
via simulations.  A comparison of the relative robustness of $\xi(r)$ and
$\Delta\xi(r)$ is presented in \S\ref{sec:results}.

As an aside we note that further low-$k$ suppression can be obtained from a
linear combination of $\xi(r)$, $\bar{\xi}(<r)$ and $\bar{\bar{\xi}}(<r)$ with
\begin{equation}
  \bar{\bar{\xi}}(<r) \equiv \frac{5}{r^5}\int_0^r x^4\,dx\ \bar{\xi}(<x)
  = \frac{5}{2}\left[ \bar{\xi}(<r) - \frac{3}{r^5}\int_0^r x^4\,dx\ \xi(x)
  \right]
\label{eqn:ddxi}
\end{equation}
for which the window function is $j_4(x)$, going as $x^4$ for $x\ll 1$.
The generalization to higher orders is straightforward but will not be
needed here.

We do not expect any of the models of $\Delta^2(k)$ to be accurate on small
scales.  Since $\bar{\xi}(<r)$ is a cumulant, using the quantity
$\Delta\xi(r)$ in place of $\xi(r)$ has traded uncertainty at large scales for
uncertainty at small scales.
However, there is a distinct advantage because we know the functional
form of the change in $\Delta\xi(r)$ at large separations due to any 
modulation of small scale power.
Assuming that the models of $\xi(r)$ are accurate at large $r$ we can write
the change in $\Delta\xi(r)$ from inaccurately modeled (or measured) small
scale power as an additive term
\begin{equation}
  \Delta\xi(r) = \Delta\xi_{\rm model}(r)+\frac{{\mathcal A}}{r^3}
  \quad {\rm with\/}\ {\mathcal A}\equiv
  3\int_0^r r'^2dr\ \left[\xi(r')-\xi_{\rm model}(r')\right] \quad .
\end{equation}
For example, if the extra power were pure shotnoise, $(k/k_1)^3$, it would
give ${\mathcal A}/r^3 = (3\pi/2)(k_1r)^{-3}$.  No matter what the small
scale physics is doing to the true value of $\xi(r)$, the integral quickly
asymptotes to a constant value for all separations greater than
${\mathcal O}(10 h^{-1}$Mpc), for which the models are a good fit.
It is useful to know the functional form of the modification;
it allows us to marginalize over the parameter 
${\mathcal A}$ when connecting the observed galaxy correlation function to the 
HOD through $b$ and $k_2$, and to the cosmology, through $\alpha$.  

Finally we point out that the generalization of $\Delta\xi(r)$ to redshift
space is straightforward and gives the same kernel for the quadrupole as
the monopole allowing easy interpretation of their ratio.
Writing the redshift space coordinate as $s$ with the cosine of the angle
to the line of sight as $\mu$ we can define
\begin{equation}
  \xi(s,\mu) \equiv \sum_{\ell=0}^{\infty} \xi_\ell(s)P_\ell(\mu)
\end{equation}
where $P_\ell(\mu)$ is a Legendre polynomials of order $\ell$ and
\begin{equation}
  \xi_\ell(s) = i^\ell \int\frac{dk}{k}\ \Delta_\ell(k)j_\ell(ks)
\end{equation}
Note that the sign of the quadrupole term in the correlation function
is opposite that in the power spectrum.
Again we can reduce the dependence of $\xi(s,\mu)$ on low-$k$ modes
by defining
\begin{equation}
  \Delta\xi(s,\mu) \equiv \bar{\xi}(<s) - \xi(s,\mu)
\end{equation}
which leads to a replacement of the $j_0$ in the monopole expression
with $j_2$ as in Eq.~(\ref{eqn:dxi_wt}).  Both $\Delta\xi_0$ and
$\Delta\xi_2$ then probe the same $k$-modes.
The hexadecapole can be isolated by using $\xi$, $\bar{\xi}$ and
$\bar{\bar{\xi}}$ as in Eq.~(\ref{eqn:ddxi}).

\section{Methodology} \label{sec:methodology}

\subsection{Fourier space}

To compute the galaxy power spectrum for the periodic, constant time, slices
we first assign our tracers to the nearest grid point of a regular, periodic,
$256^3$ Cartesian mesh and Fourier transform the resultant density
field\footnote{We get the same results by using Cloud-in-Cell interpolation
onto a $512^3$ grid.}.  The resulting $P(k)$ is corrected for the assignment
to the grid using the appropriate window function and the Poisson shot noise
is subtracted.  The power is averaged in bins spaced linearly in $k$.
The average $P(k)$ is plotted at the position of the average $k$ in each bin.
We obtain approximate error bars both by counting the number of modes in each
bin and by dividing the data into 8 disjoint octants and using the
octant-to-octant variance from the 3 different simulations -- for a total of
24 sub-samples.
The latter method underestimates the errors on large scales, while the former
neglects the mode coupling which occurs on small scales, the extent of which
depends on the galaxy sample under consideration.
When computing the power spectrum of each octant we set the remaining 7
octants to zero and correct for the windowing factor of 8 in power and the
modified shot noise when doing the FFT.
For HODs with $M_1\gg M_{\rm min}$ we find that mode counting and sub-sample
variance agree very well with
\begin{equation}
  \frac{\delta P}{P} = \sqrt{\frac{2}{N_{\rm mode}}}
  \left( 1 + \frac{1}{\bar{n}P(k)} \right)
  \qquad .
\label{eqn:modecounting}
\end{equation}
The higher order shot-noise terms \cite{MeiWhi} are subdominant for the
range of scales of interest to us.
As $M_1$ approaches $M_{\rm min}$ there is excess variance on small scales
compared to the mode counting predictions.
This is not unexpected.  For models with $M_1\simeq M_{\rm min}$ the galaxy
power spectrum goes non-linear at relatively low $k$, with the extra power
being transferred from larger scales by gravitational collapse.  The number
of modes driving the variance is thus smaller than one would estimate by
mode counting at the measured $k$.

{}From the 8 octants for each of 3 simulations we are able to compute the
covariance matrix of $P(k)$.  We find that for our chosen bin size the bins
are almost independent at the scales of interest for the oscillations.
Once the data become significantly non-linear mode coupling induces a
correlation.  For $k<0.2\,h\,{\rm Mpc}^{-1}$ the correlation coefficient is
below $5\%$, reaching 10\% by $k\simeq 0.25\,h\,{\rm Mpc}^{-1}$.
The degree of correlation between adjacent bins in Fourier space is, however,
a function of the HOD.  The more highly biased catalogs have more significant
correlations between adjacent bins on larger scales.
For catalogs with a bias of $b=3.5$ for example, adjacent bins have
correlations exceeding $0.2$ for $k>0.7\,h{\rm Mpc}^{-1}$.
However, since the correlations are so small in the $k$-range of interest
we use Eq.~(\ref{eqn:modecounting}) in the fits.

We compute our power spectra in redshift space assuming the distant observer
approximation for all outputs and use the periodicity of the simulation to
remap positions.  For the isotropically averaged spectra the power ratios
do not exactly recover the results of Ref.~\cite{Kaiser} on large scales.
Whether we should expect to reach 
the those limits on the scales relevant to baryon
oscillations remains in doubt -- see Ref.~\cite{Sco04} and references
therein for further discussion.
On small scales we are able to compute the $\Delta^2_\ell$ by direct
summation on the Cartesian $k$ grid, however on large scales we do not have
enough modes.  For this reason we perform a least squares fit for the
$\Delta^2_\ell$ up to $\ell=6$ for each of the $k$ bins.
The line-of-sight angular dependence of the power spectrum on large scales
is simple, as expected: the resulting Legendre coefficients above $\ell=4$
are small for $k\le 0.3\,h\,{\rm Mpc}^{-1}$.

%Again assuming Gaussian statistics with uncorrelated $\vec{k}$ modes we
%can compute
%\begin{equation}
%  {\rm Cov}\left[ \Delta_\ell^2(k) , \Delta_{\ell'}^2(k') \right] =
%  {\rm a\ big\ mess}
%\end{equation}

\subsection{Configuration space}

\begin{figure}
\begin{center}
\resizebox{2.7in}{!}{\includegraphics{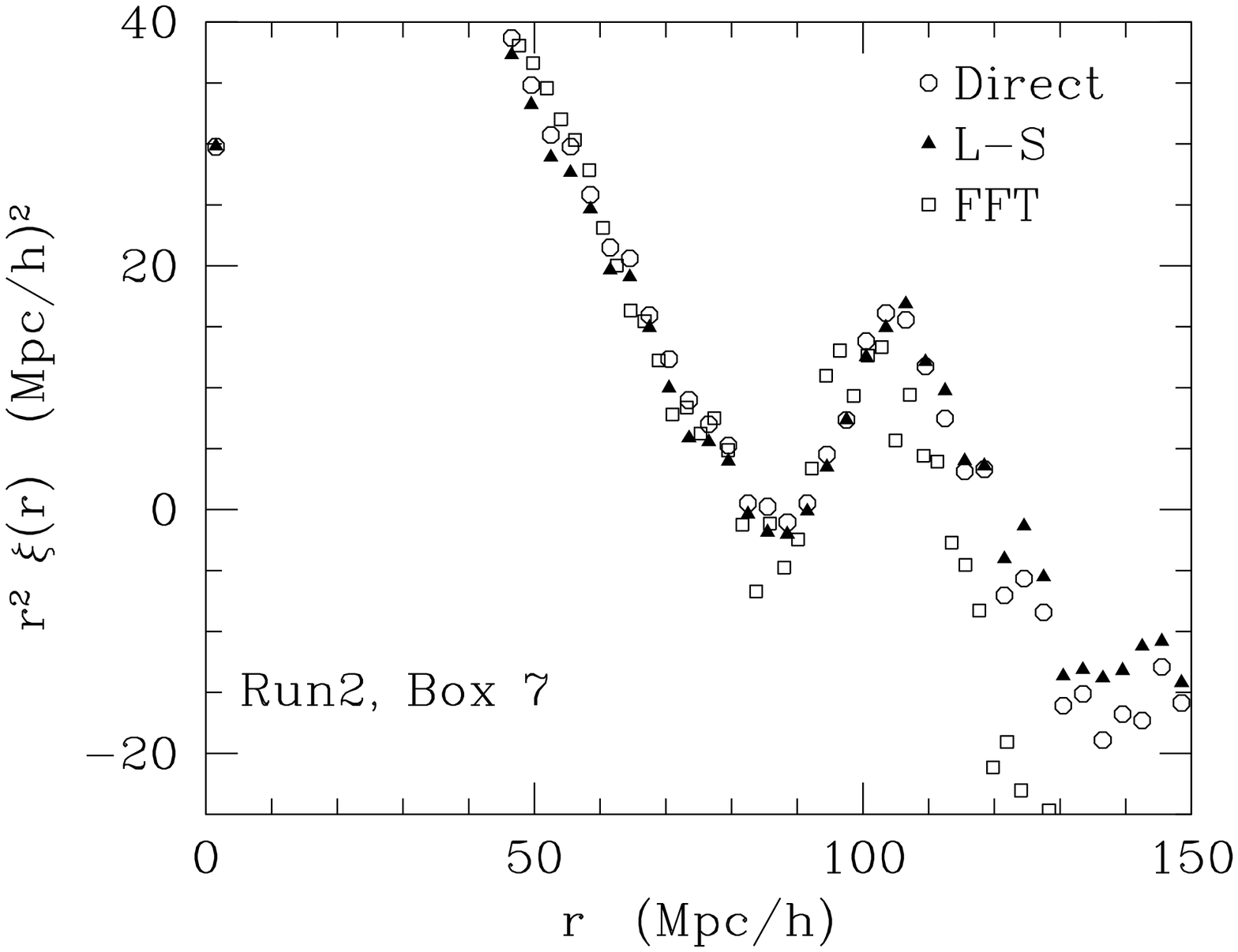}}
\resizebox{2.7in}{!}{\includegraphics{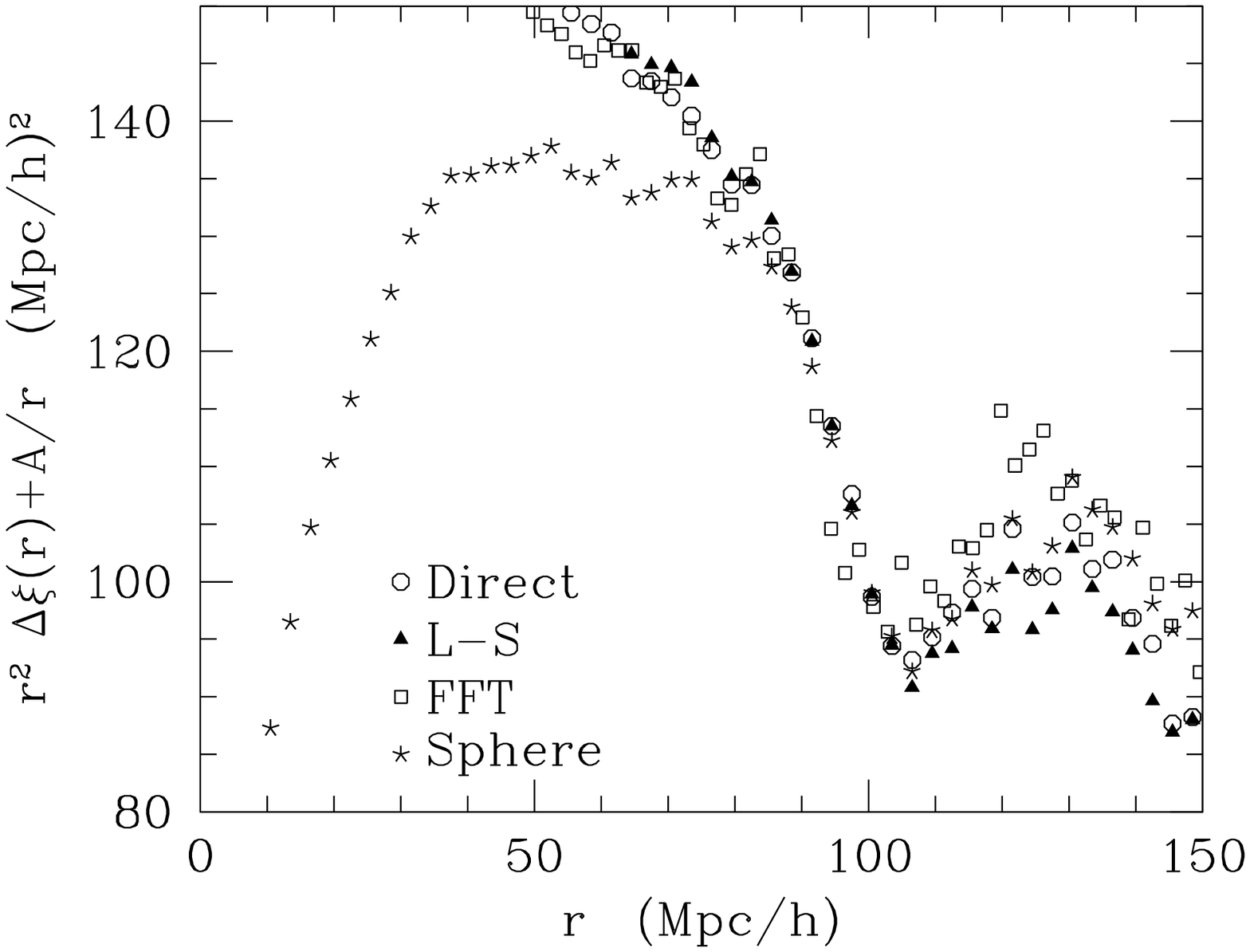}}
\end{center}
\caption{A comparison of the different ways of computing $\xi(r)$ and
$\Delta\xi(r)$ discussed in the text.  In both panels open circles represent
direct pair counting in the periodic box, filled triangles the Landy-Szalay
estimator and open squares the Fourier transform method.  For the
$\Delta\xi(r)$ plot we also show the estimate of $\Delta\xi$ where $\bar{\xi}$
is obtained from counts in spheres as the stars.  The different estimators
differ in $\xi(r)$ at small $r$, so for the $\Delta\xi$ plot we have added an
$r^{-1}$ term to make $r^2\Delta\xi=100$ at $r=100\,h^{-1}$Mpc for all
of the lines.  For $\xi(r)$ there is a noticeable shortfall in the power
estimated by FFT methods at large $r$ which is largely absent for
$\Delta\xi$.}
\label{fig:xi_cmp}
\end{figure}

We can compute $\xi(r)$ either by directly counting pairs as a function of
separation in the periodic volume, 
using FFT techniques in the periodic volume, or counting pairs and using the
Landy \& Szalay \cite{LanSza} estimator
\begin{equation}
  \xi(r) = \frac{\langle DD\rangle - 2\langle DR\rangle + \langle RR\rangle}
                  {\langle RR \rangle}
  \qquad {\rm pairs\ in\ }(r;dr)
\end{equation}
in sub-volumes with vacuum boundary conditions -- here $D$ refers to the
data, $R$ refers to a random catalog with the same selection and the angled
brackets indicate the number of pairs in a given shell $(r;dr)$.
A comparison of the different techniques is shown in Fig.~\ref{fig:xi_cmp}.
For the FFT method we use a Fourier grid with $512^3$ points.
Both CIC and NGP charge assignment \cite{HocEas} give the same answer for
$\xi(r)$ well above the grid scale.
For the Landy \& Szalay method we used a random catalog with $10\times$
as many points as the data -- increasing the number did not alter the
results -- but in computing $\langle RR\rangle$ we only searched for pairs
around the first $N_{\rm data}$ random points.
Since the $\langle DR\rangle$ point is limited to $N_{\rm data}$ points
there is no loss in accuracy from limiting the $\langle RR\rangle$ term
similarly but the cost scales as $N_{\rm data}\times N_{\rm random}$ rather
than $N_{\rm random}^2$.
In what follows we shall use the direct pair counting estimate of the
correlation function, in bins of width $3\,h^{-1}$Mpc.  We note in passing
that estimating $\xi$ from future surveys will be non-trivial as we wish
to work at very large scales with fine radial resolution.

Formally $\bar{\xi}(<r)$ can be estimated in a similar manner to $\xi(r)$
except using counts within spheres of radius $r$ rather than shells at $r$
of radius $dr$.  Unfortunately this method is prone to strong boundary effects
(in the case of the sub-volumes with vacuum boundary conditions) because the
simulation volume available to each shell in the sphere is a strong function
of radius.  Computing $\xi$ in shells and then integrating up to find
$\bar{\xi}$ is much more stable, and we do this for all of our estimators.
We tried several interpolation schemes and the results were insensitive to
our choice.  We use linear interpolation between the measured $\xi$ points
in the results below.

\begin{figure}
\begin{center}
\resizebox{5.5in}{!}{\includegraphics{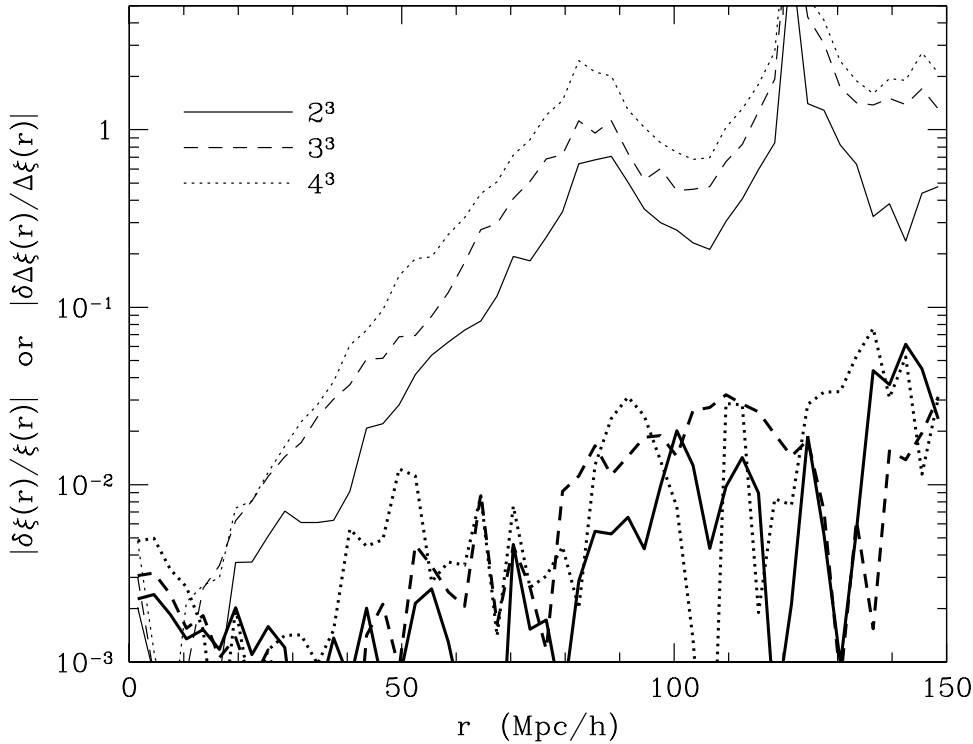}}
\end{center}
\caption{The fractional error in $\xi(r)$ (thin, upper) and $\Delta\xi(r)$
(thick, lower) computed in a smaller volume compared to the quantities in the
full simulation.  In each case we divided each simulation into $2^3$ (solid),
$3^3$ (dashed) or $4^3$ (dotted) sub-cubes and computed $\xi(r)$ or
$\Delta\xi(r)$ using the estimator of \protect\cite{LanSza} (see text).
The curves display the deviation from the average computed with periodic
boundary conditions in the full simulation. }
\label{fig:xi_vs_dxi}
\end{figure}

We expect $\Delta\xi(r)$ to be less susceptible to finite volume effects than
$\xi(r)$.  To verify this we divided each simulation into $2^3$, $3^3$ or
$4^3$ sub-cubes of side $L/2$, $L/3$ or $L/4$ and computed $\xi(r)$ and
$\Delta\xi(r)$ using the estimator of \cite{LanSza}.  We average the estimates
over sub-cubes and simulations and compare them 
in Fig.~\ref{fig:xi_vs_dxi} to the average result computed
using direct counting in 
the full box with periodic boundary conditions for each simulation. 
At larger $r$ the results become noisy, however
it is clear that while both $\xi(r)$ and $\Delta\xi(r)$ are insensitive to the
sub-division at small $r$, $\Delta\xi(r)$ suffers far less than $\xi(r)$ from
small volume effects at larger $r$.  For reference the analytically predicted
error on $\Delta\xi$, averaged over all 3 simulations for this model, is 4-6\%
over the range $100-150\,h^{-1}$Mpc, close to the difference seen in 
Fig.~\ref{fig:xi_vs_dxi} at higher $r$.

The different estimators of $\xi(r)$ differ at small radius, and for
this reason our results for $\Delta\xi(r)$ can differ by an $r^{-3}$ term at
large $r$.
As discussed earlier we also expect our models of $\Delta^2(k)$ to misestimate
$\Delta\xi(r)$ by a term ${\mathcal A}/r^3$.  For this reason we shall drop
the 1-halo contribution in $\Delta^2(k)$ in favor of an additive $r^{-3}$ term
in $\Delta\xi$ when doing the fits.  This term also soaks up any error in
computing $\Delta\xi$ arising from uncertainties in our estimate of $\xi(r)$
on small scales.  We shall marginalize over constant ${\mathcal A}$ when
fitting our theories to the observed $\Delta\xi(r)$ points.

The correlation function errors are highly correlated between different
scales.  Although we can estimate
${\rm Cov}\left[ \Delta\xi(r_1),\Delta\xi(r_2)\right]$ from sub-divisions
of the full volume we found that the covariances are so strong, and the
numerically estimated covariance matrix so noisy, that the results are
unstable.  For this reason we compute the covariance analytically, making
several approximations.
First we assume that the small-scale fluctuations contributing to
${\mathcal A}$ are independent of the large-scale fluctuations contributing
to $\Delta\xi(r)$ at large $r$.  For the large scale modes we assume Gaussian
errors on $P(k)$ with different $k$-modes being independent.
Then it is straightforward to show that \cite{Ber94,EisZal01,Coh06}
\begin{equation}
  {\rm Cov}\left[ \Delta\xi(r_1),\Delta\xi(r_2)\right] =
  \frac{(2\pi)^2}{V}\int \frac{dk}{k^4} \Delta^4(k) j_2(kr_1)j_2(kr_2)
   + \frac{\sigma_{\mathcal A}^2}{r_1^3r_2^3} + {\rm s.n.}
\end{equation}
where $V$ is the volume of the survey, in our case the simulation volume,
$\sigma_{\mathcal A}^2$ is the variance of ${\mathcal A}$ and ${\rm s.n.}$
indicates shot-noise terms.
Note that the first term scales as $V^{-1}$, indicating that all of the
errors on $\xi$ tend to zero as $V\to\infty$ but remain highly correlated.
This is to be contrasted with the errors on the power spectrum, which
remain ${\mathcal O}(1)$ for each $k$-mode but are uncorrelated between
increasingly finely spaced $k$-modes as $V\to\infty$.
The shot-noise term proportional to $1/\bar{n}$ is
\begin{equation}
  \frac{4}{V\bar{n}}\int \frac{dk}{k} \Delta^2(k) j_2(kr_1)j_2(kr_2)
\end{equation}
For $r_1>0$ and $r_2>0$ the $1/\bar{n}^2$ shot-noise terms are
\begin{equation}
  \frac{1}{2\pi V\bar{n}^2}
  \left[ \frac{3}{r_1^3}\Delta\xi(r_2) + \frac{3}{r_2^3}\Delta\xi(r_1) +
  \frac{3}{r_>^3}\Delta\xi(r_<) + \frac{1+\xi(r_1)}{r_1^2}\delta(r_1-r_2)\right]
\end{equation}
where $r_<$ the lesser of $r_1$ and $r_2$ and $r_>$ is the greater.
The $\delta$-function in the last term is rendered finite when we estimate
$\xi$ in bins of finite width.  For small bins we can replace $\delta(r_1-r_2)$
with the inverse of the bin width.  Since $\xi\ll 1$ at large scales this
gives $r_1^{-2}\Delta r^{-1}$ along the diagonal.
Finally the $1/\bar{n}^3$ term is simply
\begin{equation}
  \left(\frac{1}{4\pi}\right)^2 \frac{1}{V\bar{n}^3}
  \ \frac{3}{r_1^3}\frac{3}{r_2^3}
\end{equation}
While we show in \S\ref{sec:results} that making this Gaussian assumption
does not bias the results for $\alpha$, in future it would be better to use
a large set of mock catalogs to estimate the covariance.  From our limited
numbers of realizations, and using the non-linearly processed Gaussian fields,
we found that the shape and amplitude of the analytic expression were within
${\mathcal O}(25\%)$ of the numerical results for scales near
$100\,h^{-1}$Mpc.  The agreement at smaller scales was considerably worse.

\begin{figure}
\begin{center}
\resizebox{2.7in}{!}{\includegraphics{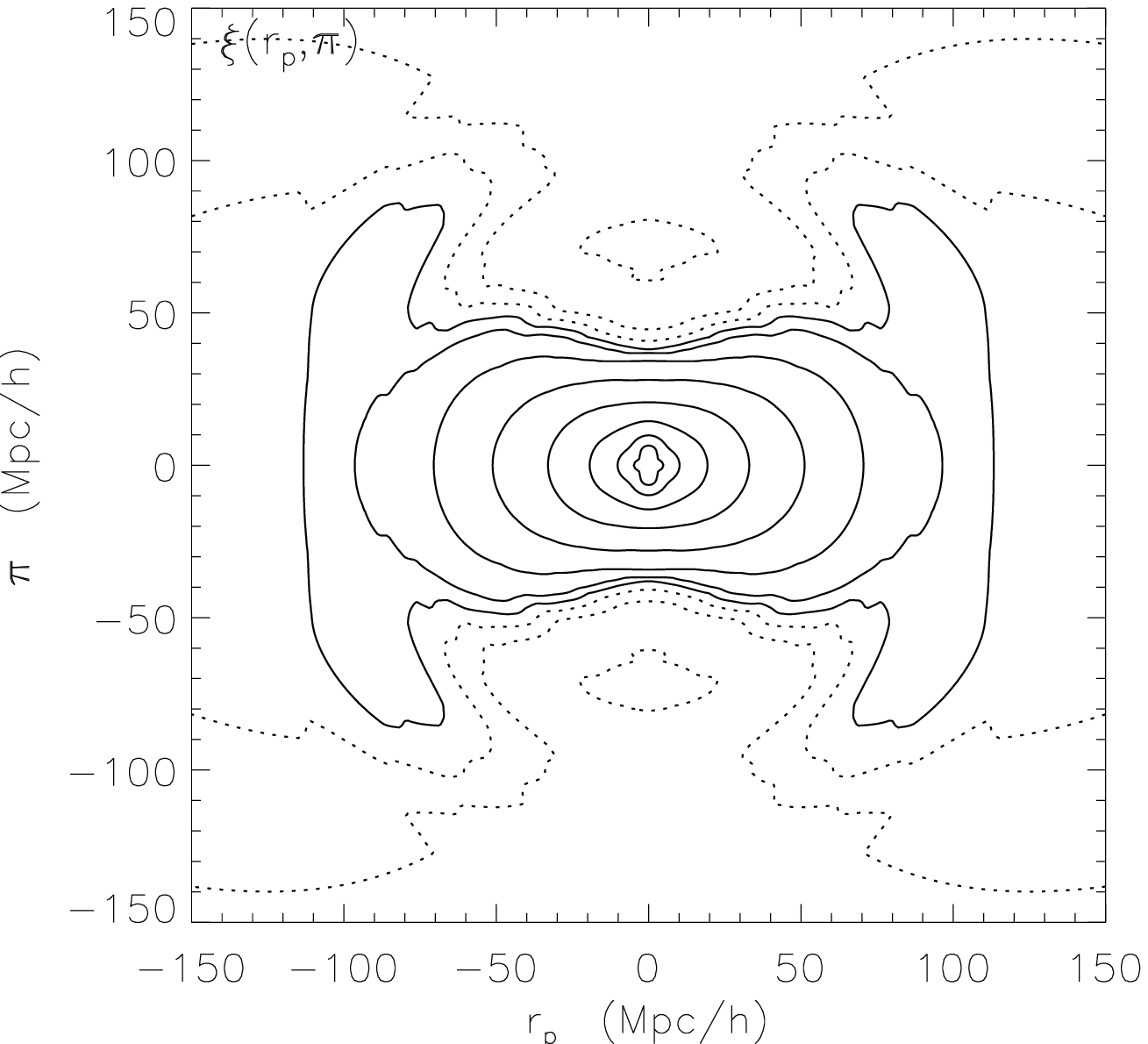}}
\resizebox{2.7in}{!}{\includegraphics{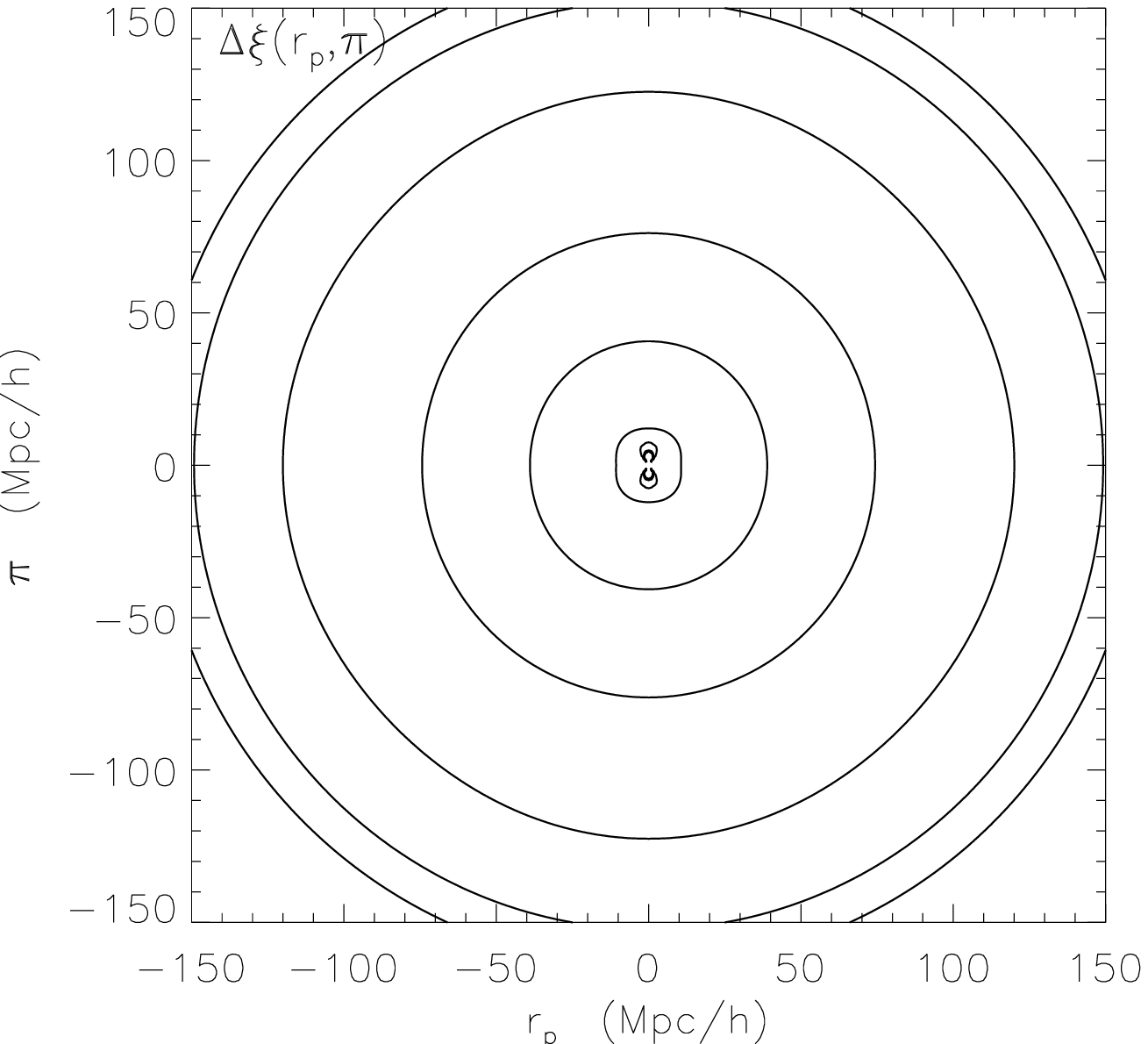}}
\end{center}
\caption{Contours of $\xi(r_p,\pi)$ (left) and $\Delta\xi(r_p,\pi)$ (right)
for one of our $b\simeq 2$ catalogs.  Contours are equally spaced in log,
and dotted lines indicate negative values.  Here $r_p$ measures separations
across the line-of-sight and $\pi$ along it.}
\label{fig:xicont}
\end{figure}

In configuration space there are no peculiar issues with estimating the $\mu$
dependence of $\xi(s,\mu)$ on the scales of interest.  We bin $\xi(s,\mu)$
in 15 bins in $|\mu|$ and then sum to get $\xi_\ell(s)$.  As 
was found for $P(k)$, the
modes beyond the quadrupole are small at large $s$.  Performing a least
squares fit to $\ell=0$, 2 and 4 yields the same result to better than a
percent for $\ell=0$, 2 and a few percent for the very small $\ell=4$ mode.
We find in general that $\Delta\xi$ is much closer to spherical than $\xi$
due to the large contribution from $\bar{\xi}(s)$.  For example from
$75\,h^{-1}$Mpc to $125\,h^{-1}$Mpc, $\Delta\xi_2/\Delta\xi_0$ falls from
approximately $0.05$ to $\simeq 0.02$ for our fiducial $b\simeq 2$ catalog.
The contours of both $\xi$ and $\Delta\xi$ are shown in
Figure \ref{fig:xicont}.

\section{Results}\label{sec:results}

\subsection{Dark matter}

\begin{figure}
\begin{center}
\resizebox{5.5in}{!}{\includegraphics{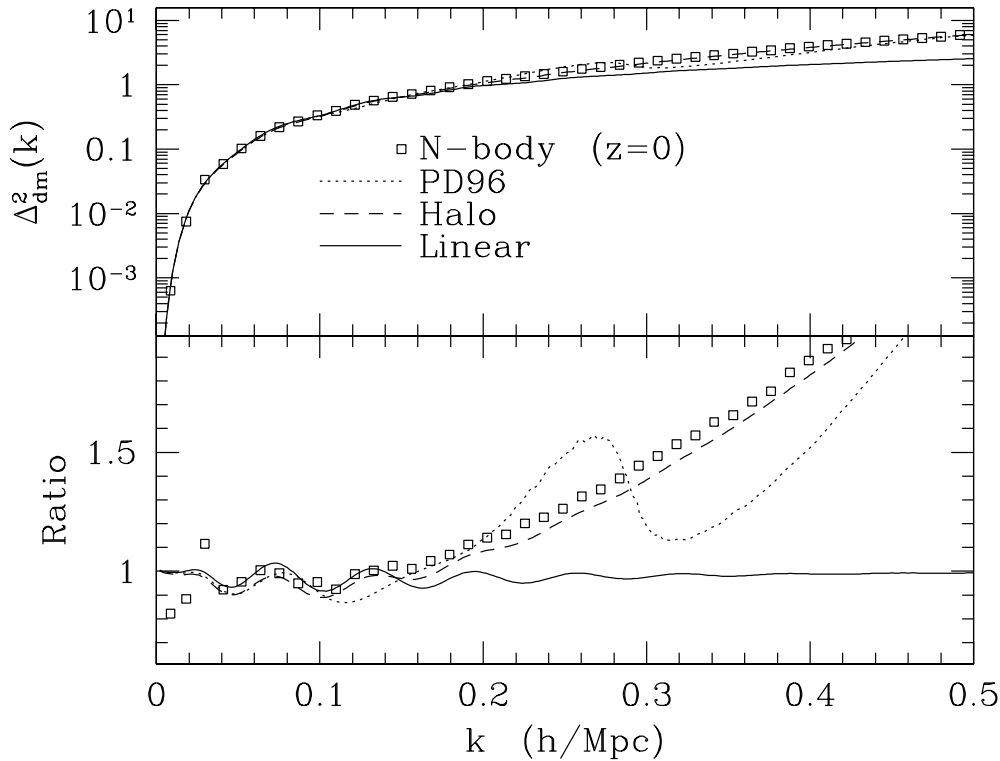}}
\end{center}
\caption{The real space power spectrum of the dark matter at $z=0$ for
one of our simulations along with two ans\"{a}tze commonly used in the
literature (see text).  The lower panel shows the ratio of the fits and
N-body points to the smooth spectrum of \protect\cite{EisHu99}.}
\label{fig:pk_fit}
\end{figure}

We begin by presenting\footnote{We thank Wayne Hu for suggesting we
make this point explicitly.} the power spectrum of the dark matter, in
real space, at the present epoch ($z=0$) from one of the simulations:
Fig.~\ref{fig:pk_fit}.
Also on the plot we show the results of two ans\"{a}tze for non-linear
spectra, that of Peacock \& Dodds \cite{PD96} based on an idea by
Hamilton et al.~\cite{HKLM}, and another
based on halo model ideas \cite{HaloFit}.
The former is seen to be a bad approximation as it implicitly assumes
that there exists a $1-1$ mapping between linear and non-linear power.
While not an issue for smoothly varying spectra, this causes problems when
the spectrum contains features such as the baryon oscillations.  In reality
mode coupling erases features, whereas the mapping procedure enhances them.
We could reduce some of the discrepancy by using a broad band measure of
the slope in the fitting function, but the underlying problem still remains.
The halo-model based methods perform better in this regard, as expected
\cite{Sel00}, since they model the non-linear power with an integral over
the linear theory power spectrum.  None of the fitting formulae approach
percent level accuracy in the non-linear regime.

\subsection{Galaxies}

\begin{figure}
\begin{center}
\resizebox{5.5in}{!}{\includegraphics{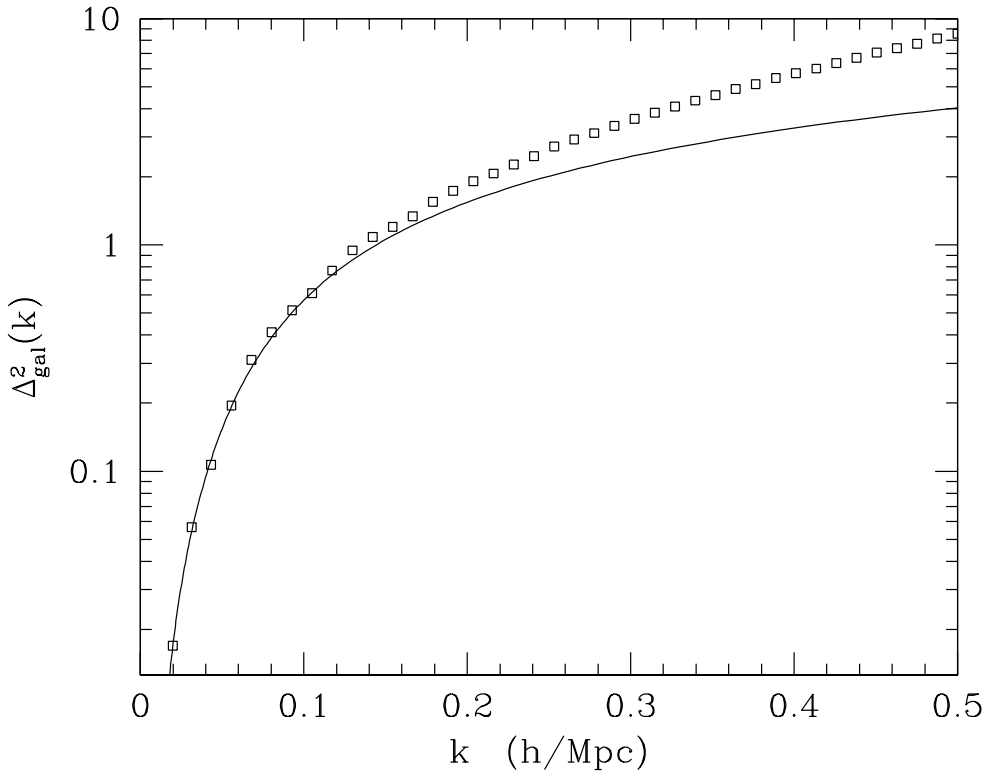}}
\end{center}
\caption{The real space power spectrum of one of our mock galaxy catalogs
with $b\simeq 2$ at $z=1$.  The solid line shows the predictions of linear
theory, multiplied by $b^2$.}
\label{fig:pk_run7}
\end{figure}

Now we turn to the mock galaxy catalogs.
We show the results at $z=1$ for one of our HOD prescriptions, with
$\bar{n}=10^{-3}\,h^3{\rm Mpc}^{-3}$ and $b\simeq 2$, in Fig.~\ref{fig:pk_run7}
along with the predictions of linear theory multiplied by $b^2$.
The power is biased on large scales and shows a clear excess on small scales. 
We shall now try to fit this behavior using the models described in
\S\ref{sec:models}.  Our results for the sound horizon parameter $\alpha$
are reported in Table \ref{tab:alpha}.

We have tried several methods for performing these fits.  We fit to both
$P(k)$ and $\Delta\xi(r)$.  For the power spectrum we use errors from
Eq.~(\ref{eqn:modecounting}), since they agree with the errors estimated from
the dispersion among the octants.
For the correlation function we use the analytic expression of
\S\ref{sec:methodology}.
The multi-dimensional fitting was done using the Levenberg-Marquardt algorithm
\cite{NumRec}.  We experimented with several implementations and found good
convergence with both analytic and numerically computed derivatives.  From
these fits we also obtain an estimate of the parameter 
covariance matrix from the
curvature of the likelihood around the best fit.
To test the Gaussianity of the likelihood surface we also ran Markov-Chain
Monte-Carlo fits (see e.g.~\cite{MCMC} for an introduction) for the power
spectra for each model.
We provide comparisons of each of these methods for the different models below.

We begin with the linear bias model (Eq.~\ref{eqn:linbias}) fit to
$P(k)$ over the range $0.02<k<k_{\rm max}\simeq 0.15\,h\,{\rm Mpc}^{-1}$.
We impose the lower $k$ cutoff so that there are enough independent modes
in the bin that the Rayleigh distribution is close to symmetric.
In performing this and subsequent fits we convolve the known $k$-space
sampling with the model at each bin to sample the theory in the same manner
as the data, although this has only a small effect on our results.
We found that the strength of the small-scale excess is a strong function of
the HOD parameters\footnote{In principle the form of the HOD for the sample
being selected can be fit from the smaller scale clustering data that we are
not analyzing here.}.  The sense is as expected -- larger small-scale excess
for models with $M_1$ closer to $M_{\rm min}$ or steeper power-law slopes
for the satellite term in $N(M)$.
When $M_1\simeq M_{\rm min}$ multiple galaxies reside in a single halo,
enhancing small-scale power.  Because the 1- and 2-halo terms shift by
different amounts in going from dark matter to galaxies, the bias is both
larger and more scale-dependent \cite{ToyModel} with the galaxy 1-halo term
dominating at larger characteristic scales than it does in the dark matter.
When $M_1\gg M_{\rm min}$ the HOD has a long `plateau', with many halos
containing only a single galaxy, and the galaxy bias is smaller and less
scale-dependent (recall we hold $\bar{n}$ fixed in our models).
For the former class of HODs, the 1-halo term can become significant at
a level of interest to BAO surveys at $k$ comparable to or even below the
standard choice for $k_{\rm max}$.  This results in biases in the sound
horizon of up to $10\%$, many times the formal fitting error.
Reducing $k_{\rm max}$ further to compensate for these cases is problematic
because -- as has been pointed out before \cite{Fisher} -- the information
content of a BAO experiment is very sensitive to this cutoff.
Thus safely fitting the power spectrum in this way requires throwing away
a significant amount of useful information.
To get around this one must accurately model the smaller scale or 1-halo
effects.

\begin{figure}
\begin{center}
\resizebox{2.7in}{!}{\includegraphics{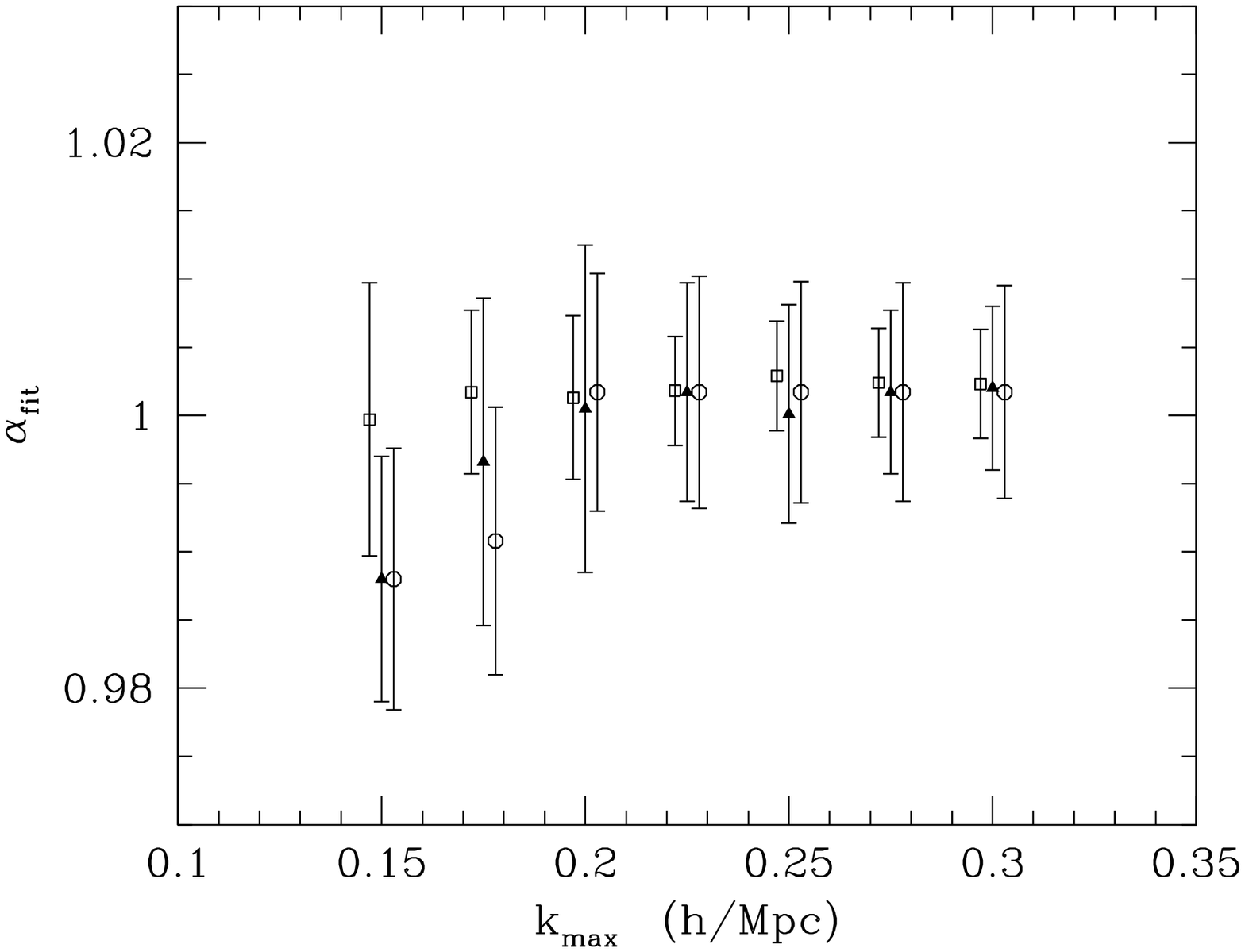}}
\resizebox{2.7in}{!}{\includegraphics{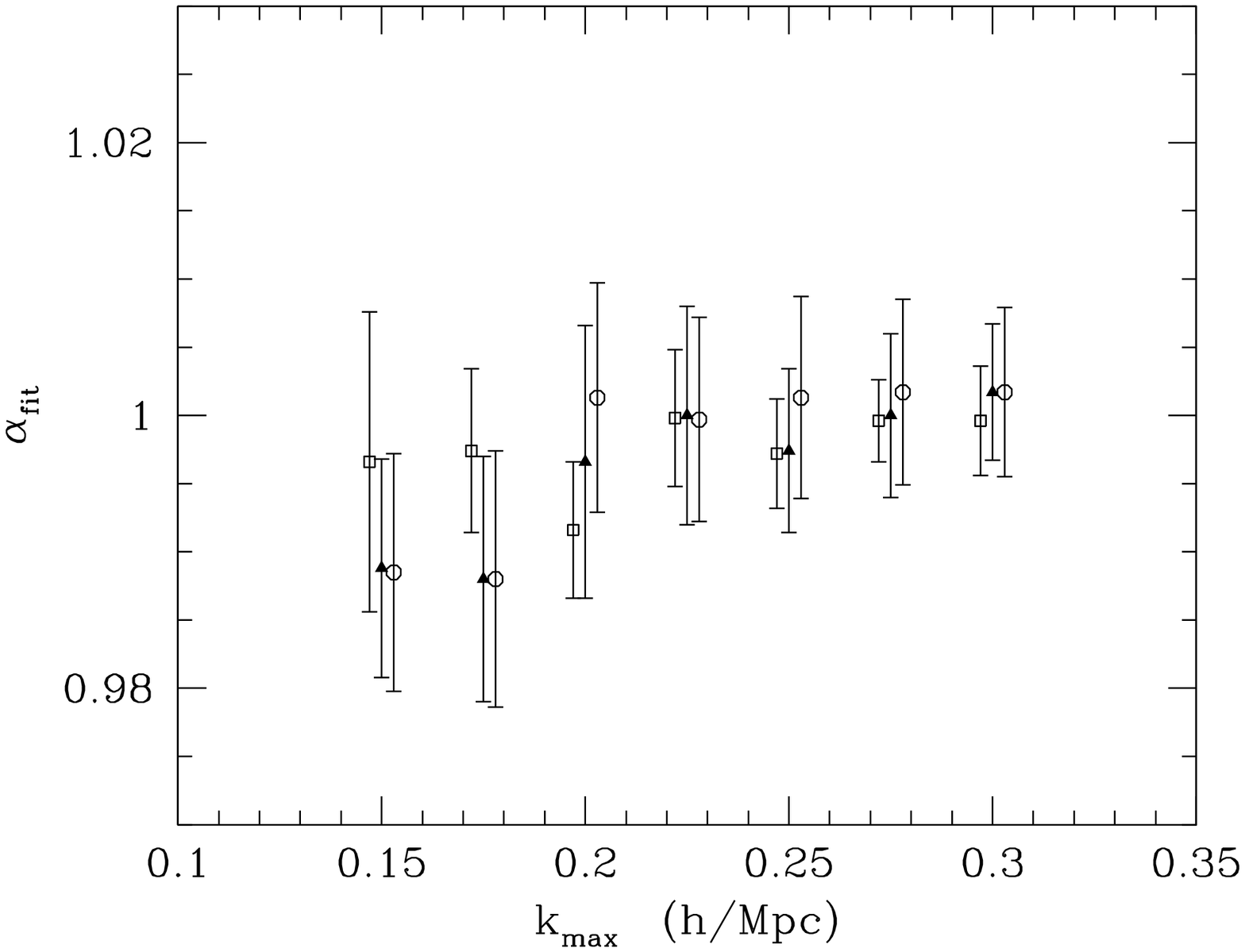}}
\end{center}
\caption{The sound horizon vs.~$k_{\rm max}$ for two different HODs:
(left)  $\log_{10}M_{\rm min}=12.8$ and $\log_{10}M_1=13.0$ and
(right) $\log_{10}M_{\rm min}=12.7$ and $\log_{10}M_1=13.5$.
Open squares are Eq.~(\protect\ref{eqn:lrg}),
solid triangles Eq.~(\protect\ref{eqn:hmi})
and open circles Eq.~(\protect\ref{eqn:esw}).
Points have been offset horizontally for clarity.}
\label{fig:kmax}
\end{figure}

For the catalogs with $\bar{n}=10^{-3}\,h^3{\rm Mpc}^{-3}$ and the models of
Eqs.~(\ref{eqn:lrg}-\ref{eqn:esw}) we find we can safely fit up to
$k_{\rm max}\simeq 0.3\,h{\rm Mpc}^{-1}$ before we notice a bias coming from
incorrectly modeled small-scale physics.  We show the best fit $\alpha$, and
the fit error, as a function of $k_{\rm max}$ for several models in
Fig.~\ref{fig:kmax}.  Note that the points are not independent because the
information content is cumulative in $k_{\rm max}$.  Beyond the range plotted
the bias in $\alpha$ becomes significant.
The fitting form of Eq.~(\ref{eqn:blagla}) does not fare as well.
To be conservative in what follows we choose
$k_{\rm max}=0.15\,h{\rm Mpc}^{-1}$ for the fits to Eq.~(\ref{eqn:blagla})
and $k_{\rm max}=0.25\,h{\rm Mpc}^{-1}$ for the other models.  Beyond this
$k_{\rm max}$ our assumption of Gaussian uncorrelated errors becomes
increasingly suspect.

For the lower-density catalogs with $\bar{n}=10^{-3.5}\,h^3{\rm Mpc}^{-3}$,
we find similar results when fitting the same models. Adopting a small-scale
cutoff of $k_{\rm max}=0.25\,h{\rm Mpc}^{-1}$ does not tend to bias the fitting
results. Significant differences between the fit results at differing
densities only begin to appear when dealing with highly biased populations.

Fitting the form of Eq.~(\ref{eqn:blagla}) to the data in the range
$0.02\le k\le 0.15\,h\,{\rm Mpc}^{-1}$ and assuming the `correct'
$\Delta_{\rm ref}$ we find the best fitting $k_A$.  Dividing the `true'
$k_A\simeq 0.06\,h{\rm Mpc}^{-1}$ by the best fit we find
$\alpha\simeq 1.00$ for essentially all of our HOD models, with a mild
trend for higher $\alpha$ (about $1\sigma\approx 1\%$) for the less biased
samples with the smaller satellite fractions.  If we extend $k_{\rm max}$
beyond $0.15\,h\,{\rm Mpc}^{-1}$ we find $\alpha$ increases and becomes
inconsistent with unity.
Using all 3 of our runs, for a total survey volume of
$3.8\times 10^9\,h^{-3}{\rm Gpc}^3$, we are able to constrain $\alpha$ to
${\mathcal O}(1\%)$ for our catalogs -- consistent with the expectations of
simple error propagation.

The model of Eq.~(\ref{eqn:lrg}) provides a reasonable fit to the data over
the relevant $k$ range: $0.02\le k\le 0.25\,h\,{\rm Mpc}^{-1}$.
The Markov chains converge very rapidly for this model, and the best fits
for $b$ and $\alpha$ were insensitive to the precise value of $a$ chosen.
This suggests that the first few elements of a Taylor series expansion in
the multiplicative bias would work as well.
The marginalized distributions for $b$, $Q$ and $\alpha$ are reasonably well
fit by Gaussians, with $\alpha=0.996\pm0.005$ for our $b\simeq 2$ catalog for
example.  The best fit provided by the Levenberg-Marquardt algorithm for this
model is $\alpha=0.998\pm0.004$.
Compared to the halo model forms (Eqs.~\ref{eqn:hmi}, \ref{eqn:esw}) the best
fitting models with Eq.~(\ref{eqn:lrg}) have less intermediate scale power.
To compensate, the value for $b$ tends to be a few percent higher
($b\simeq 2.17$ for this model).
This also leads to a slightly higher $\chi^2$ for the fit.
Comparing the galaxy $P(k)$ to that of the dark matter we find $b(k)$ rises
from $2.05$ near the fundamental mode to $2.2$ at $k=0.2\,h{\rm Mpc}^{-1}$
and $2.3$ at $k=0.4\,h{\rm Mpc}^{-1}$.  The best fit bias is thus
${\mathcal O}(10\%)$ higher than the $k\to 0$ value.

\begin{figure}
\begin{center}
\resizebox{2.7in}{!}{\includegraphics{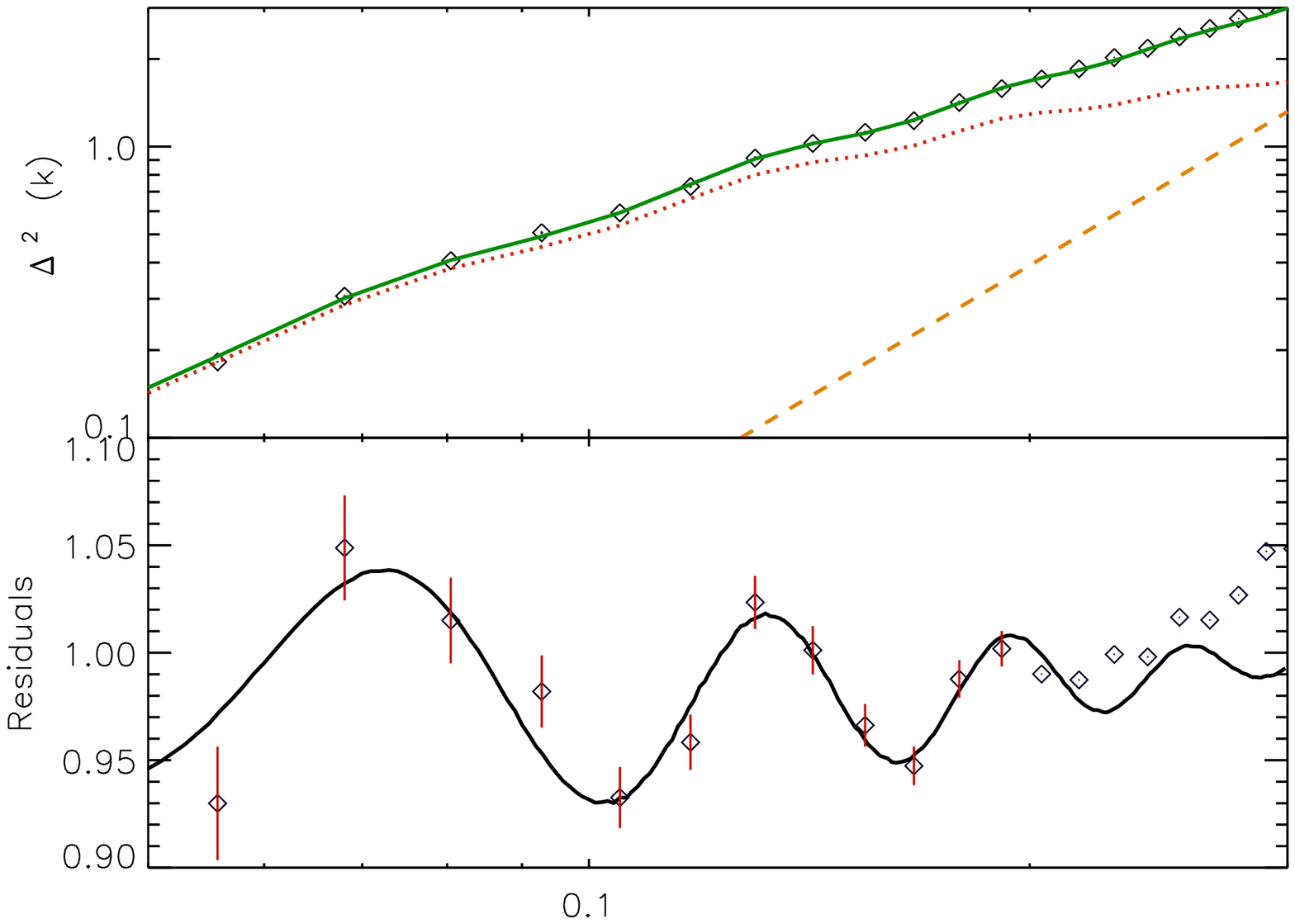}}
\resizebox{2.7in}{!}{\includegraphics{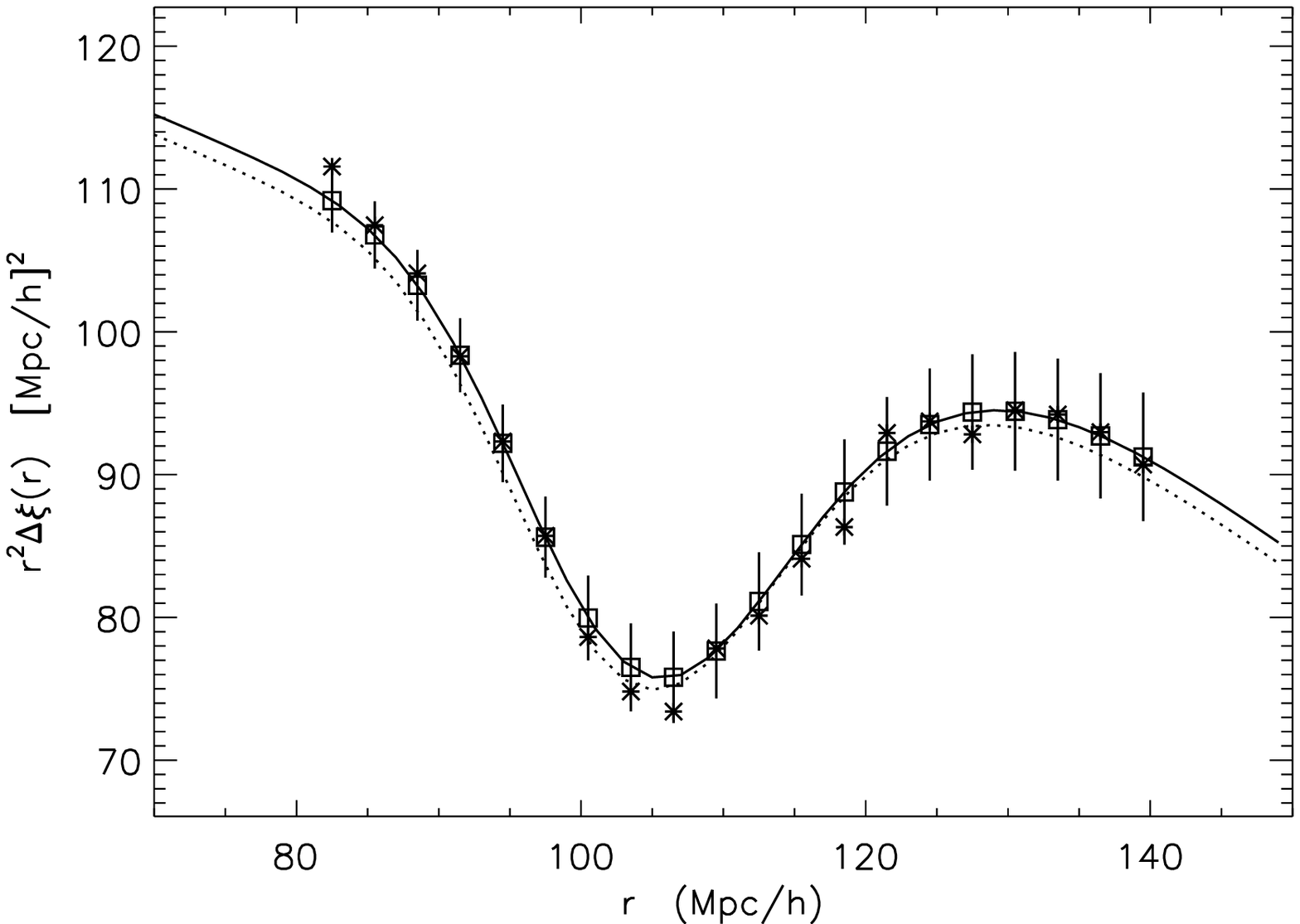}}
\end{center}
\caption{(Left) The real space power spectrum, $\Delta^2(k)$, for one of our
HOD models along with the halo model based fit of Eq.~(\protect\ref{eqn:hmi}).
The points are derived from the mock galaxy catalog, the dotted and dashed
lines show the biased linear theory and shot-noise terms for the best fit
with the solid line being their sum.  The lower panel shows the ratio of
the theory and points to the smooth spectrum of \protect\cite{EisHu99} along
with the fit residuals.
(Right) The correlation function, $\Delta\xi$, along with the best fit
(solid) and the best fit to the power spectrum (dotted).  Note the errors
on $\Delta\xi$ have been computed analytically and so have been placed on
the best fit curve rather than the data (stars).  The errors are highly
covariant.}
\label{fig:pk_real}
\end{figure}

Figure \ref{fig:pk_real} shows the residuals around the fit to $P(k)$
when using Eq.~(\ref{eqn:hmi}) and the catalog with $b\simeq 2$.
As is evident from the figure, Eq.~(\ref{eqn:hmi}) is a good representation
of the real-space power spectrum.
Comparisons between the fit results for different HODs show that $k_1$ (the
1-halo parameter) and $b$ (the large-scale bias) vary as described in
\cite{ToyModel} with little scatter.
The Markov chains show the marginalized $\alpha$ distribution is consistent
with unity, within $1\sigma$, for all of the catalogs.
We also show in Fig.~\ref{fig:pk_real} the best fit model using the
$\Delta\xi(r)$ data for $80\,h^{-1}{\rm Mpc}<r<140\,h^{-1}$Mpc.
The agreement between the $P(k)$ and $\Delta\xi(r)$ best fits is within
$1\sigma$ for the bias, $k_2$ and the sound horizon.
It is difficult to meaningfully compare the values of $k_1$, but the fit
prefers a negligible 1-halo term for models with large $M_1$ as expected.
For the model of Eq.~(\ref{eqn:hmi}) the predicted shape $\Delta\xi$ falls
below the data for smaller $r$.  For this reason the values of the parameters
returned are sensitive to the range of $r$ chosen in the fit.
In general the fits to $\Delta\xi$ have slightly worse $\chi^2$ values than
for $P(k)$, but the shape of the $\chi^2$ surface for $\alpha$ is similar.

\begin{figure}
\begin{center}
\resizebox{5.5in}{!}{\includegraphics{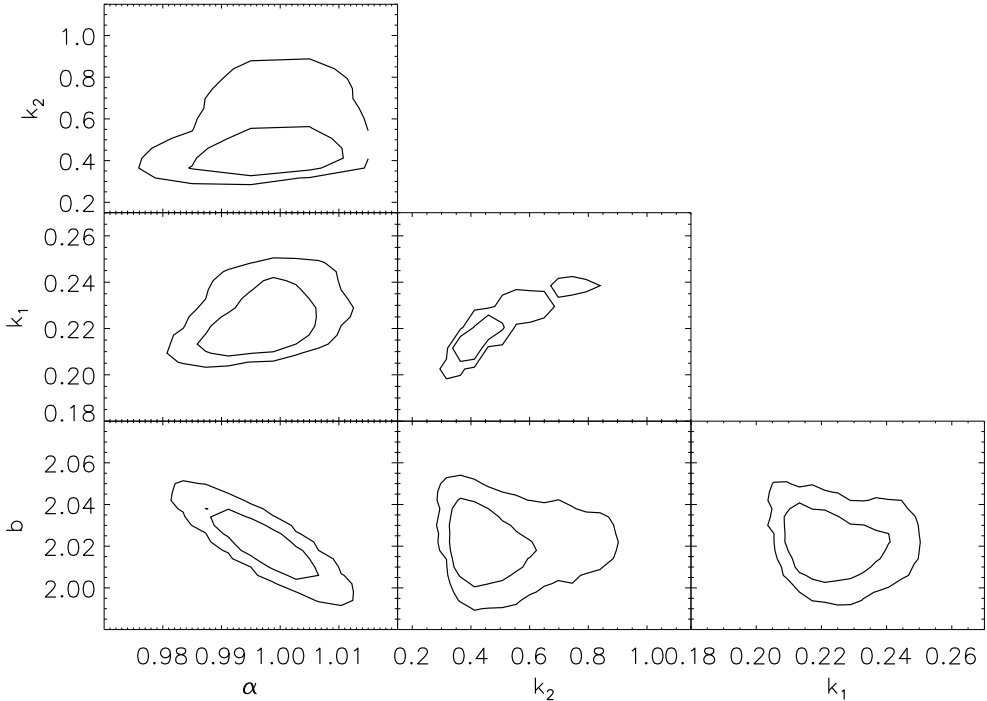}}
\end{center}
\caption{The marginalized errors in each of the fit parameters of
Eq.~(\protect\ref{eqn:hmi}) for one of our catalogs with $b\simeq 2$.
Contours represent $68\%$ and $95\%$ enclosed probability.}
\label{fig:MargErr}
\end{figure}

We show marginalized error contours (68 and 95\% enclosed probability) for
the 4 parameters of our $P(k)$ fit to the same catalog in
Fig.~\ref{fig:MargErr}.  These regions are derived from our Markov chains;
the error ellipses derived from the curvature of the likelihood at maximum
are similar except for $k_1$ where the distribution is not well fit by a
Gaussian.
Except for the large-scale bias, $b$, the contours show little degeneracy
between the cosmology, as parameterized by $\alpha$, and the properties of
the galaxy sample.  The $b-\alpha$ degeneracy is more pronounced for
galaxy populations with larger 1-halo terms.
As the small-scale power excess is increased, and the acoustic features are
more washed out, reducing the sound horizon (increasing $\alpha$) becomes
a way of shifting power from large scales to small, and is thus difficult to
distinguish from a change in the amplitude of the 1-halo term.
The MCMC derived marginalized likelihoods are close to Gaussian for $b$,
$k_2$ and $\alpha$, but there is a significant tail to low $k_1$.

The form of Eq.~(\ref{eqn:esw}) fits $\Delta\xi$ better to lower $r$ than
Eq.~(\ref{eqn:hmi}).  As for Eq.~(\ref{eqn:hmi}) there is a slight trend
for HOD models with higher $M_1$ to have lower $\alpha$, but again the best
fits are within $1\%$ of unity for all of our catalogs.
The marginalized parameter distributions from the MCMC run on the $P(k)$
data are close to Gaussian except for $k_1$ and $k_2$, with the latter
poorly constrained for this model.  For example, the marginalized distribution
for $\alpha$ for the $b\simeq 2$ data plotted above is well fit by
$\alpha=1.000\pm0.007$.

\begin{figure}
\begin{center}
\resizebox{5.5in}{!}{\includegraphics{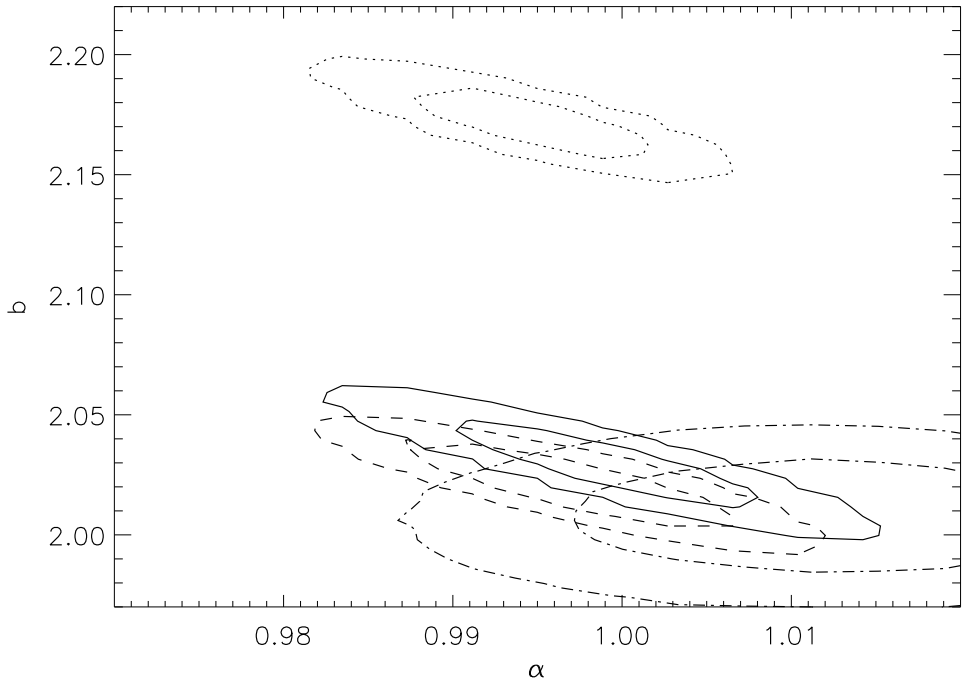}}
\end{center}
\caption{The marginalized errors in $b$ and $\alpha$ computed from the
Markov chains fit to our $b\simeq 2$ mock catalog for several of the
models discussed above:
Eq.~(\protect\ref{eqn:blagla}) dot-dashed, Eq.~(\protect\ref{eqn:lrg}) dotted,
Eq.~(\protect\ref{eqn:hmi}) dashed and Eq.~(\protect\ref{eqn:esw}) solid.
Contours represent $68\%$ and $95\%$ enclosed probability.}
\label{fig:mcmc_all_ba}
\end{figure}

Finally we compare in Fig.~\ref{fig:mcmc_all_ba} the marginalized errors in
$b$ and $\alpha$ from the Markov chain fits to the catalog in
Fig.~\ref{fig:pk_run7}.  The value of $b$ preferred by Eq.~(\ref{eqn:lrg})
is slightly high, as discussed above.

\begin{table}
\begin{center}
\begin{tabular}{ccccc}
  $\log_{10}M_{\rm min}$ & $\log_{10}M_1$  &
  Eq.~(\ref{eqn:lrg})    &
  Eq.~(\ref{eqn:hmi})    &
  Eq.~(\ref{eqn:esw})   \\
  12.8 & 13.0 & $1.003\pm0.005$ & $0.998\pm0.007$ & $0.999\pm0.008$ \\
  12.7 & 13.5 & $0.996\pm0.005$ & $0.998\pm0.006$ & $1.000\pm0.007$ \\
  12.6 & 14.5 & $0.991\pm0.005$ & $0.998\pm0.006$ & $0.999\pm0.006$
\end{tabular}
\end{center}
\caption{Results of fitting our catalogs to the various functional forms for
three typical choices of HOD parameters at
$\bar{n}=10^{-3.0}\,h^3\,{\rm Mpc}^{-3}$.  We quote the parameters of a
Gaussian fit to the marginalized $\alpha$ distribution from our Markov chains.}
\label{tab:alpha}
\end{table}

\subsection{Redshift space clustering} \label{sec:redresult}

The results so far have been in real space, as appropriate for surveys
measuring projected clustering statistics.  We now turn to measures in
redshift space.
We begin by discussing the angle-averaged power spectrum\footnote{We neglect
the contribution of $\Delta_2^2$ to the covariance of $\Delta_0^2$.  For
our models, on the scales of relevance, this is a good approximation.}
and correlation function.  The constraints on $\alpha$ should thus be
interpreted as an average of the transverse and line-of-sight distances, or
approximately a shift in $(D^2H^{-1})^{1/3}$.

Since the form of Eq.~(\ref{eqn:blagla}) did not perform well for the
real space tests we begin by considering Eq.~(\ref{eqn:lrg}).
In terms of the angle-averaged clustering pattern in redshift space we
have found that for $k<0.25\,h{\rm Mpc}^{-1}$ the exponential, Lorentzian
or Gaussian streaming models fit the small-scale downturn to sufficient
accuracy for our purposes, though they each prefer different values of
$\sigma$.  The goodness of fit is best for the Gaussian suppression.
If we multiply Eq.~(\ref{eqn:lrg}) by a Gaussian suppression and a
$1+2\beta/3+\beta^2/5$ prefactor (here $\beta\equiv\Omega^{0.6}/b$) we
obtain good fits to most of the mock catalogs.
Taking our $b\simeq 2$ catalog for example, the marginalized real space
constraint is $\alpha=0.996\pm0.005$ while the redshift space constraint
is $\alpha=0.997\pm0.008$.  For the bias we obtain $b=2.17\pm0.01$
vs.~$b=2.14\pm0.02$ while the small-scale suppression is only poorly
constrained.
The difference in $b$ reflects the inapplicability of the large-scale
enhancement factor.
With this $k_{\rm max}$ there is a very slight bias in $\alpha$ for the
highest $b$ HOD model.
The results are essentially unchanged if we use a Lorentzian suppression
model, but if we use exponential suppression there is a 1\% downward bias
on $\alpha$ even for the $b\simeq 2$ catalog and the goodness of fit is
noticeably worse.

For the forms Eqs.~(\ref{eqn:hmi}, \ref{eqn:esw}) we get good fits for either
a Gaussian or exponential suppression.  In each case the marginalized $\alpha$
distribution is consistent with unity with an uncertainty of just under 1\%.
The form of Eq.~(\ref{eqn:esw}) does slightly better, as in the real space
case.

\begin{figure}
\begin{center}
\resizebox{2.7in}{!}{\includegraphics{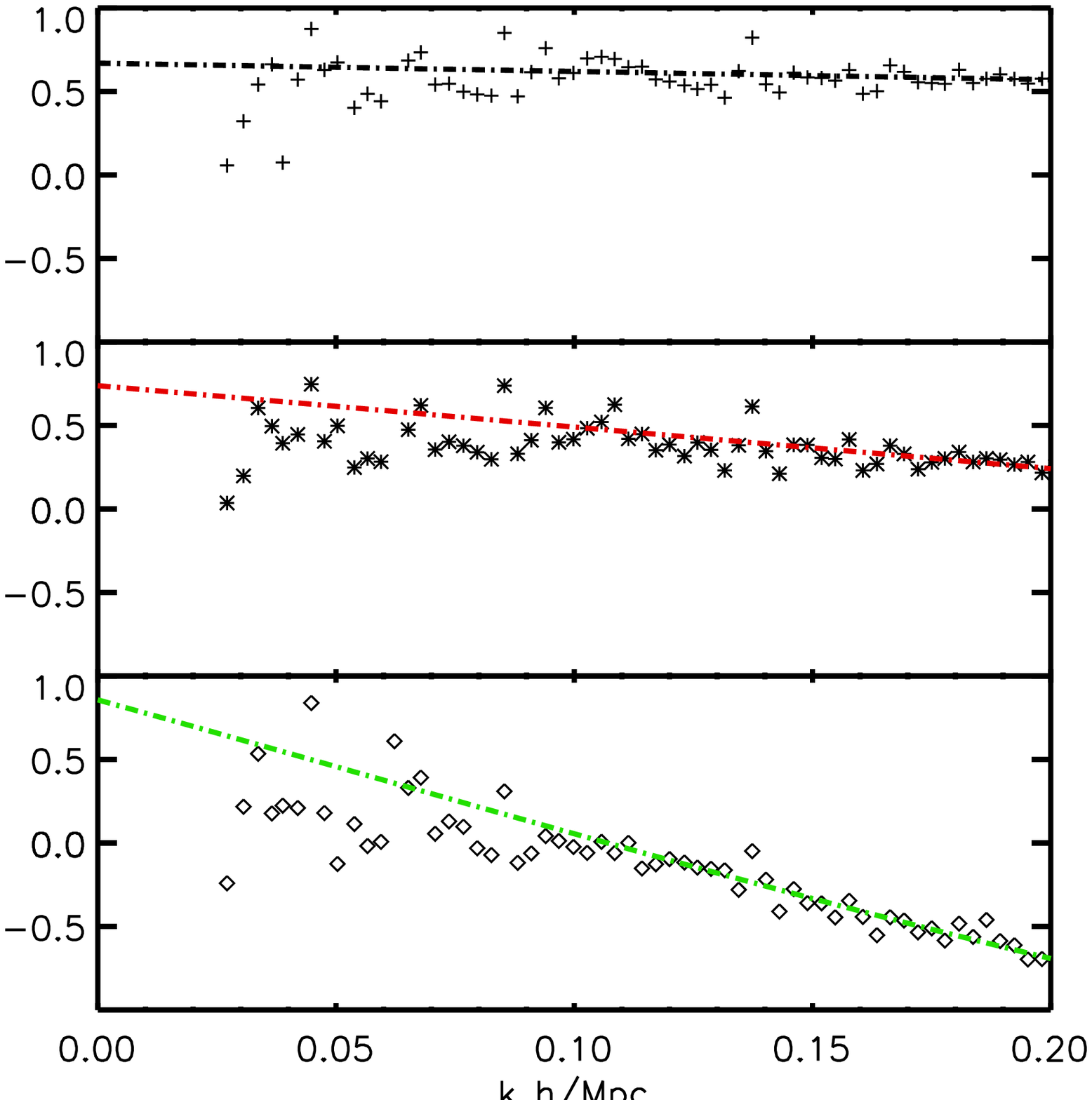}}
\resizebox{2.7in}{!}{\includegraphics{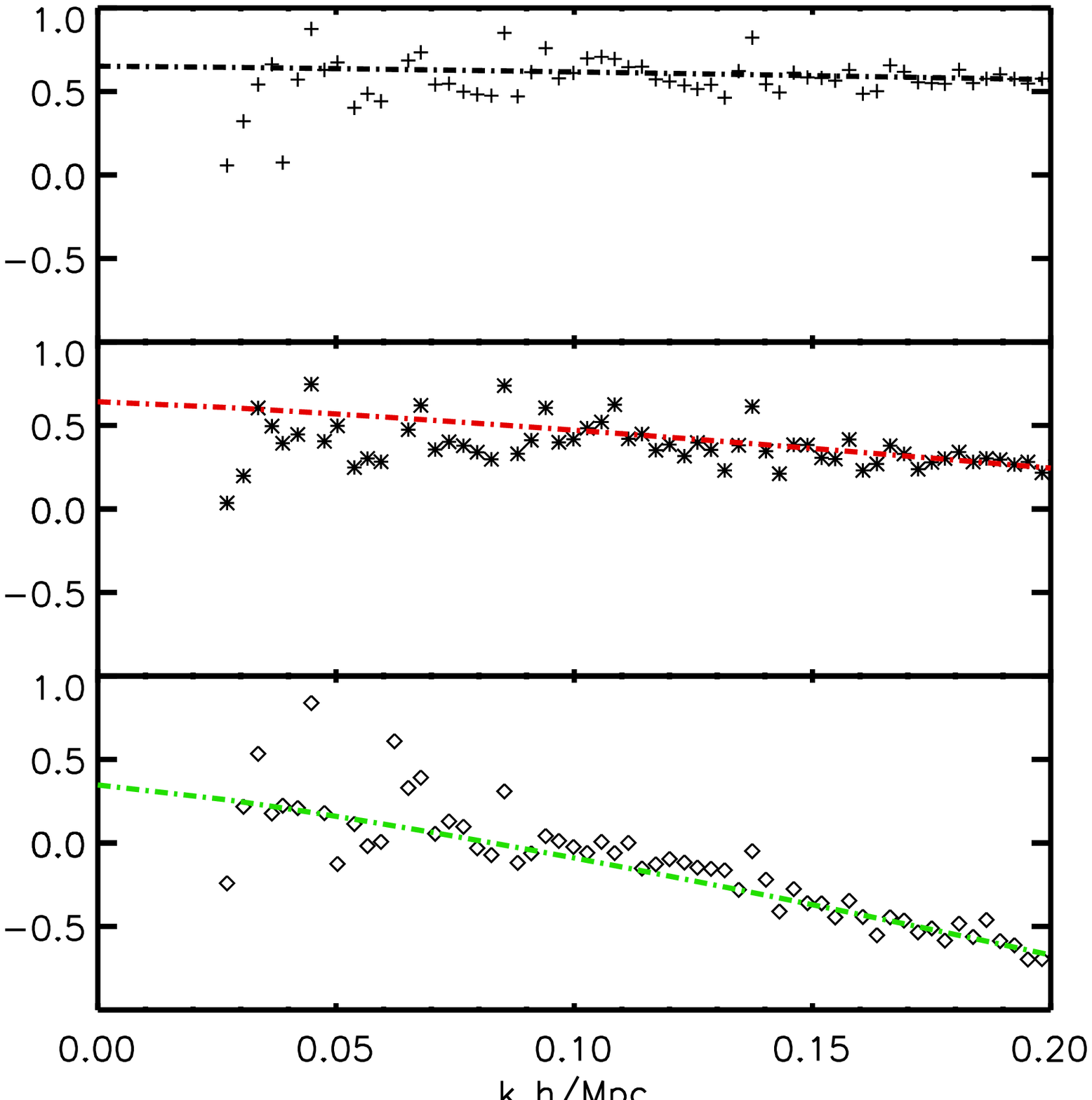}}
\end{center}
\caption{(Left) The quadrupole-to-monopole ratio vs.~scale for a number of
our models at $z=1$ along with streaming model fits (see text).  The streaming
model tends to overestimate $Q/M$ at low-$k$ for the more biased models
(lower panels).  (Right)  The same models with the fit of
Ref.~\protect\cite{HatCol}.  In our catalogs the high $k$ suppression closely
tracks the mean galaxy weighted halo mass (see text).}
\label{fig:q2m}
\end{figure}

Now let us consider the angle dependence of the clustering.
Figure \ref{fig:xicont} shows the anisotropy in configuration space,
i.e.~for $\xi$ and $\Delta\xi$ for one of our models.
Figure \ref{fig:q2m} shows the quadrupole-to-monopole ratio in Fourier
space for a number of our HOD models with
$\bar{n}=10^{-3}\,h^3\,{\rm Mpc}^{-3}$.
As can be seen in the figure, we find a strong HOD dependence to the redshift
space clustering of galaxies.  In the limiting case of halos-only
(i.e.~models with no satellite contribution) there is little small-scale
suppression even on scales as small as $1.5\,h\,{\rm Mpc}^{-1}$.

\begin{figure}
\begin{center}
\resizebox{2.7in}{!}{\includegraphics{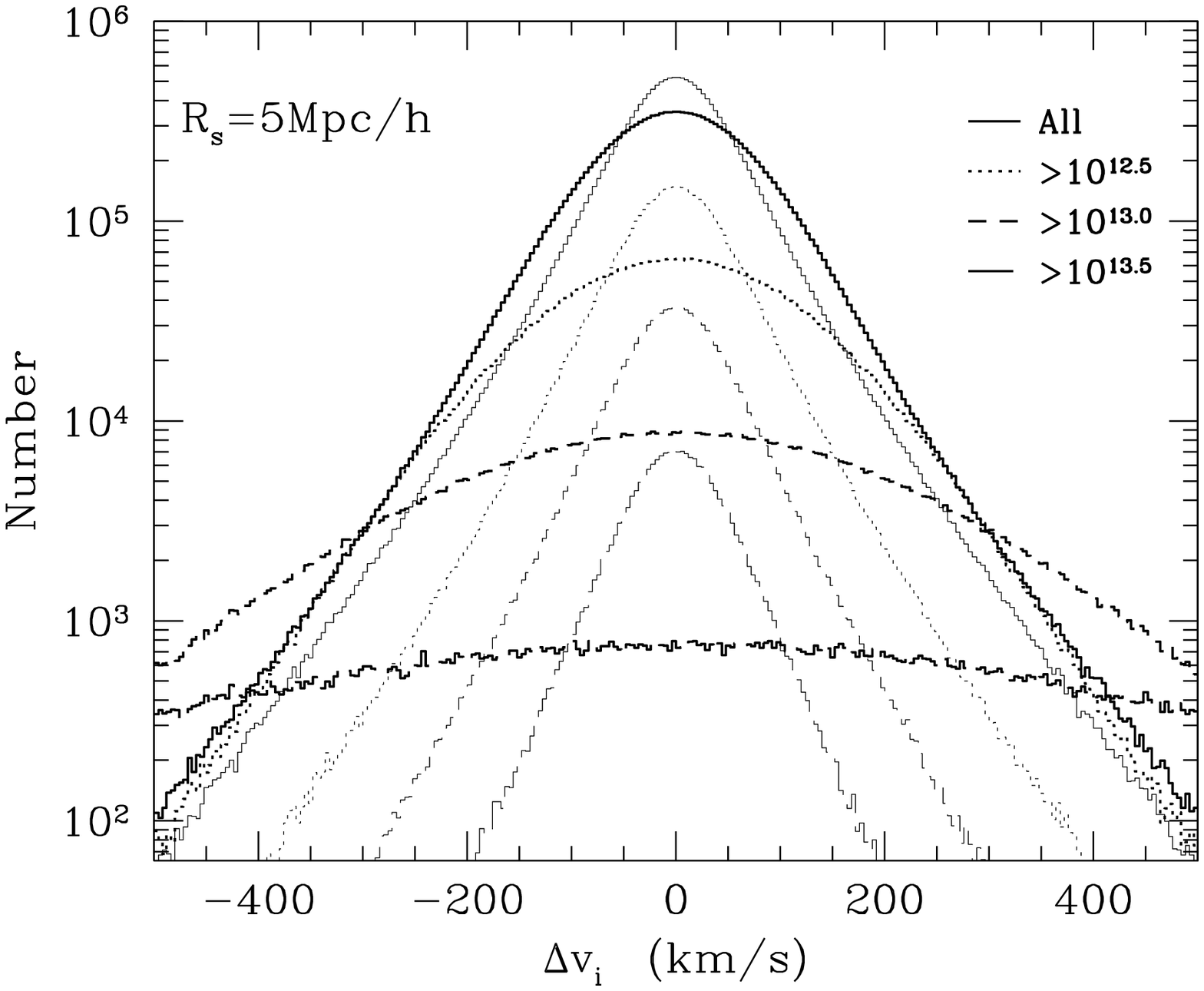}}
\resizebox{2.7in}{!}{\includegraphics{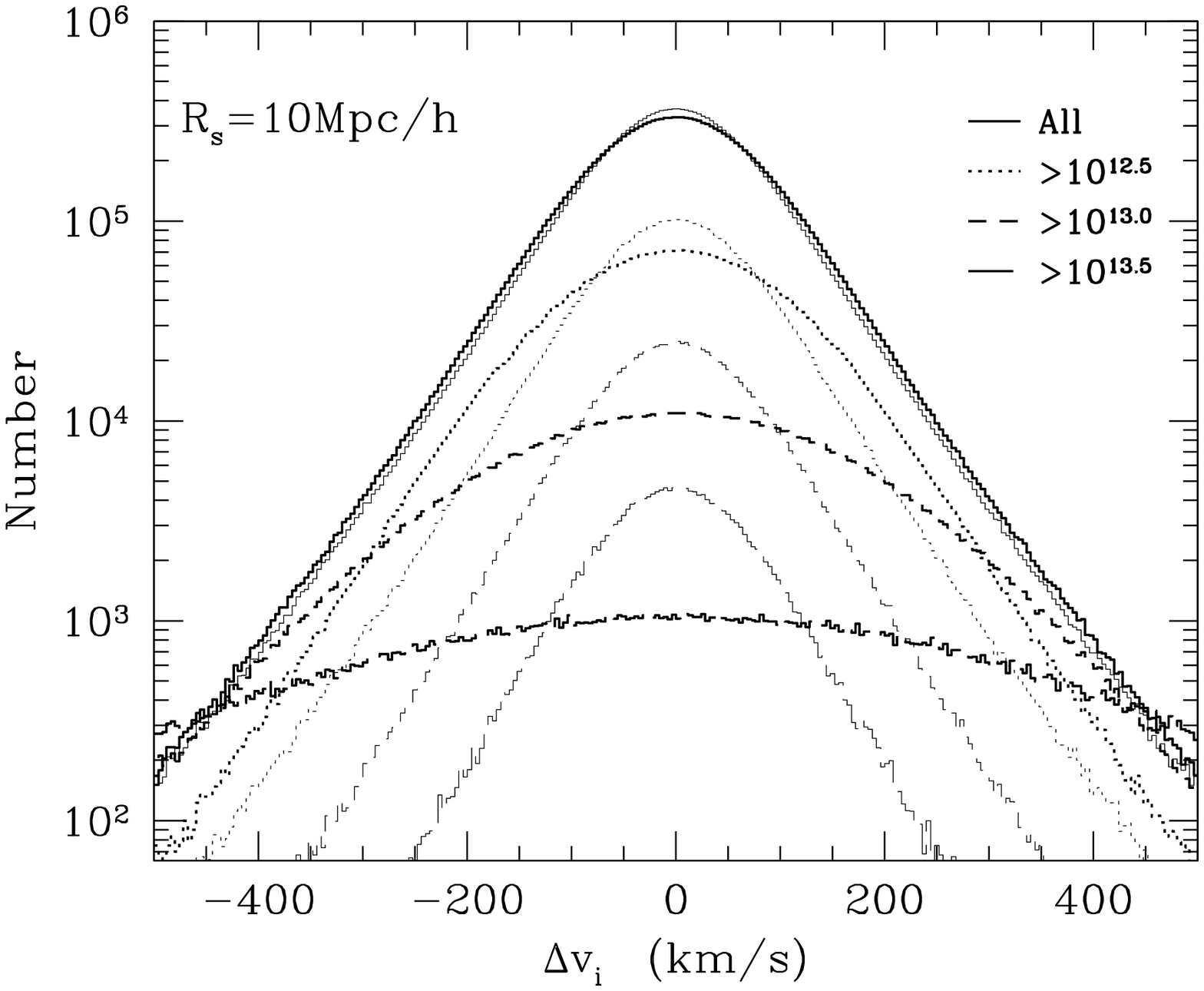}}
\end{center}
\caption{The true halo velocity in the simulations compared to the linear
theory prediction from a smoothed density field with Gaussian width
$5\,h^{-1}$Mpc (left) or $10\,h^{-1}$Mpc (right).  The thick lines show
the histogram of velocity differences in each component when the density
field is estimated from the halo positions, the thin lines when the density
field is estimated directly from the dark matter particles in the simulation.}
\label{fig:velcmp}
\end{figure}

Further insight into this behavior comes from considering the deviation
between the true halo velocity in the simulation and that which would be
predicted using linear theory from the (smoothed) density field in the
simulation at $z=1$ (see also \cite{SheDia01,CroEfs94,Col00,PadBau} for
similar calculations).
This is shown in Fig.~\ref{fig:velcmp} for two smoothing scales.
When the density field is estimated from the dark matter particles in the
simulation the rms deviation in each component of $v$ is between 50 and
$80\,$km/s with smaller deviations for higher mass halos and smaller smoothing
scales.  If the velocity is estimated from the halo catalog (weighting all
halos equally) then the rms rises to between 90 and $260\,$km/s.  The increase
in the rms comes from the neglect of the mass contributed by the lower
mass halos, the fact that more massive halos tend to live in denser
environments and the assumption of equal weights per halo
\cite{SheDia01,Col00}.
The fact that the true velocities differ only slightly (the equivalent of a
few Mpc) from those predicted by linear theory suggests we should see little
small-scale suppression in the redshift space halo power spectrum on the
scale of interest.

The other extreme case -- the strongest small-scale deviation from the
Kaiser approximation -- occurs where the mean galaxy-weighted halo mass
and the satellite fraction are at their highest.  For these HOD models
the galaxies preferentially sample the highest peculiar velocities.
The other HODs occupy a continuum between these cases that is well explained
by the fraction of satellite galaxies and the mean galaxy-weighted halo mass.
Overplotted in Fig.~\ref{fig:q2m} are the fits to the functional form
derived from a streaming model with exponential small-scale
suppression.  We also attempted to fit streaming models with Gaussian
and Lorentzian cutoffs, but these were generally poorer fits to the
data than the exponential cutoff. The streaming model fits shown are
acceptable fits to the simulation quadrupole-to-monopole ratios only
when the latter varies slowly with $k$; our more highly biased models
(lower panels) are not well-represented.
In Fig.~\ref{fig:q2m} we also compare our results with the fitting formula
of Eq.~(\ref{eqn:hatcol}).  This is quite a good fit to the simulations
over the range of interesting scales.

Additional constraints on power spectrum model parameters can be gleaned from
the redshift-space distortions.  One option is to use the linear theory
parameter $\beta$ as a simultaneous constraint on the cosmology and the galaxy
bias (recall the large-scale bias, $b$, was the parameter most degenerate
with $\alpha$, see Fig.~\ref{fig:MargErr}).
We examined this possibility, and found that even on the largest scales
in the simulations, the best fits to the multipole moments did not approach
the linear theory value of $\beta$ with enough precision for this to be useful.

One property of the angular dependence of the redshift-space distortions that
is well constrained is the scale of the quadrupole zero-crossing -- represented
by $k_{\rm nl}$ in the empirical fit of \cite{HatCol}, and by $\sigma$ in the
streaming model.  This quantity is in fact very tightly correlated with the
choice of HOD parameters.  In this sense, at least one independent constraint
on the relevant galaxy physics derived from redshift-space distortions is
fairly insensitive to the choice between these models.

We do not consider using the angular dependence of the redshift space
distortions to fit separately for the line-of-sight and transverse
distance scales.  Before further considering redshift space distortions
we want to include the (light-cone) evolution of clustering.

\section{Reconstruction} \label{sec:reconstruction}

The common lore is that surveys targeting galaxy populations with a large
1-halo term (small values of $k_1$), or surveys at low $z$ will have fewer
`useful' modes than high redshift surveys or surveys of objects which singly
occupy their halos.
Recent work \cite{ESSS06} has suggested that it may be possible to partially
mitigate these effects and reconstruct the baryon oscillation signal despite
the corrosion due to non-linear collapse, even for surveys at low redshifts.

\begin{figure}
\resizebox{2.7in}{!}{\includegraphics{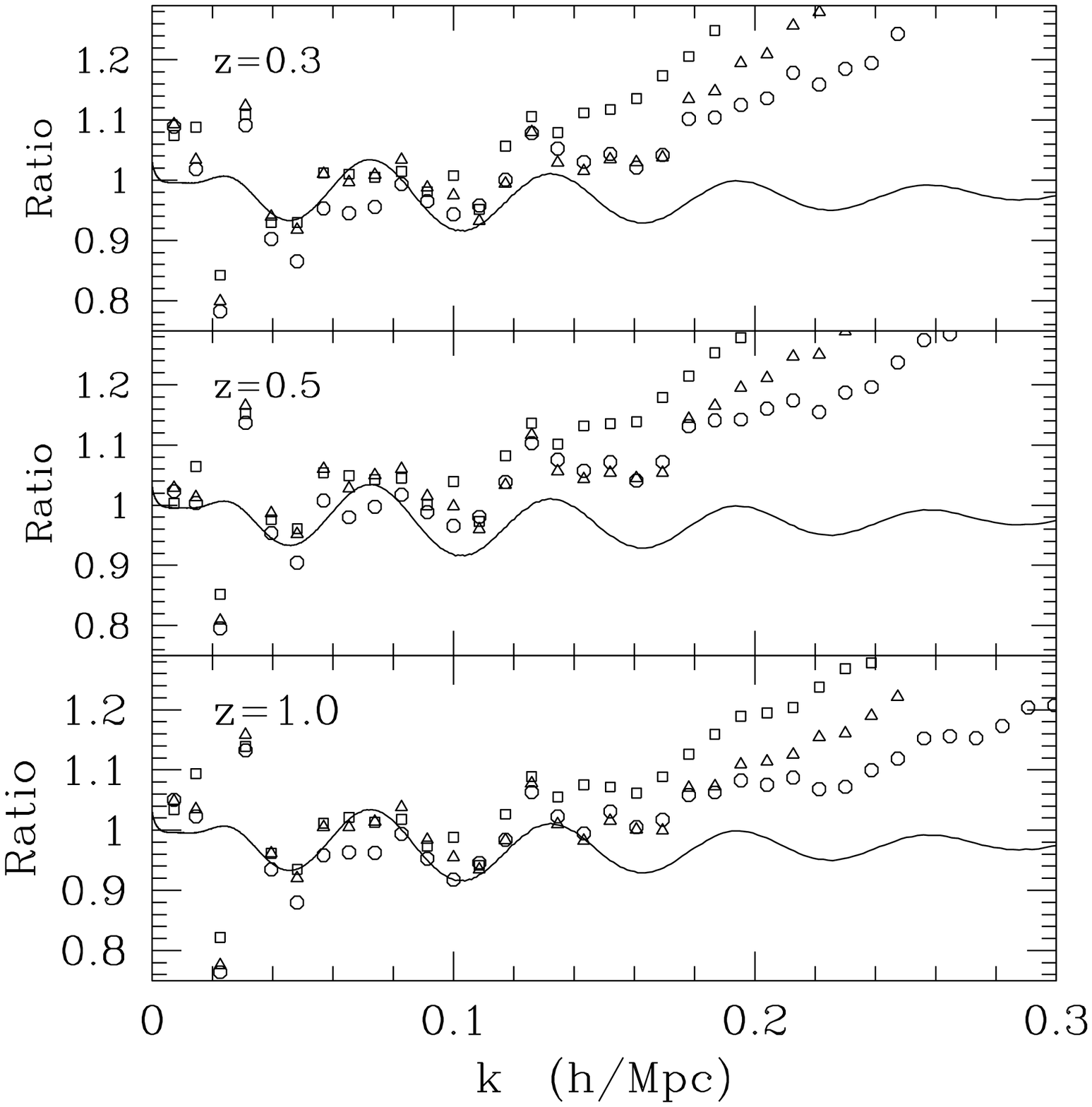}}
\resizebox{2.7in}{!}{\includegraphics{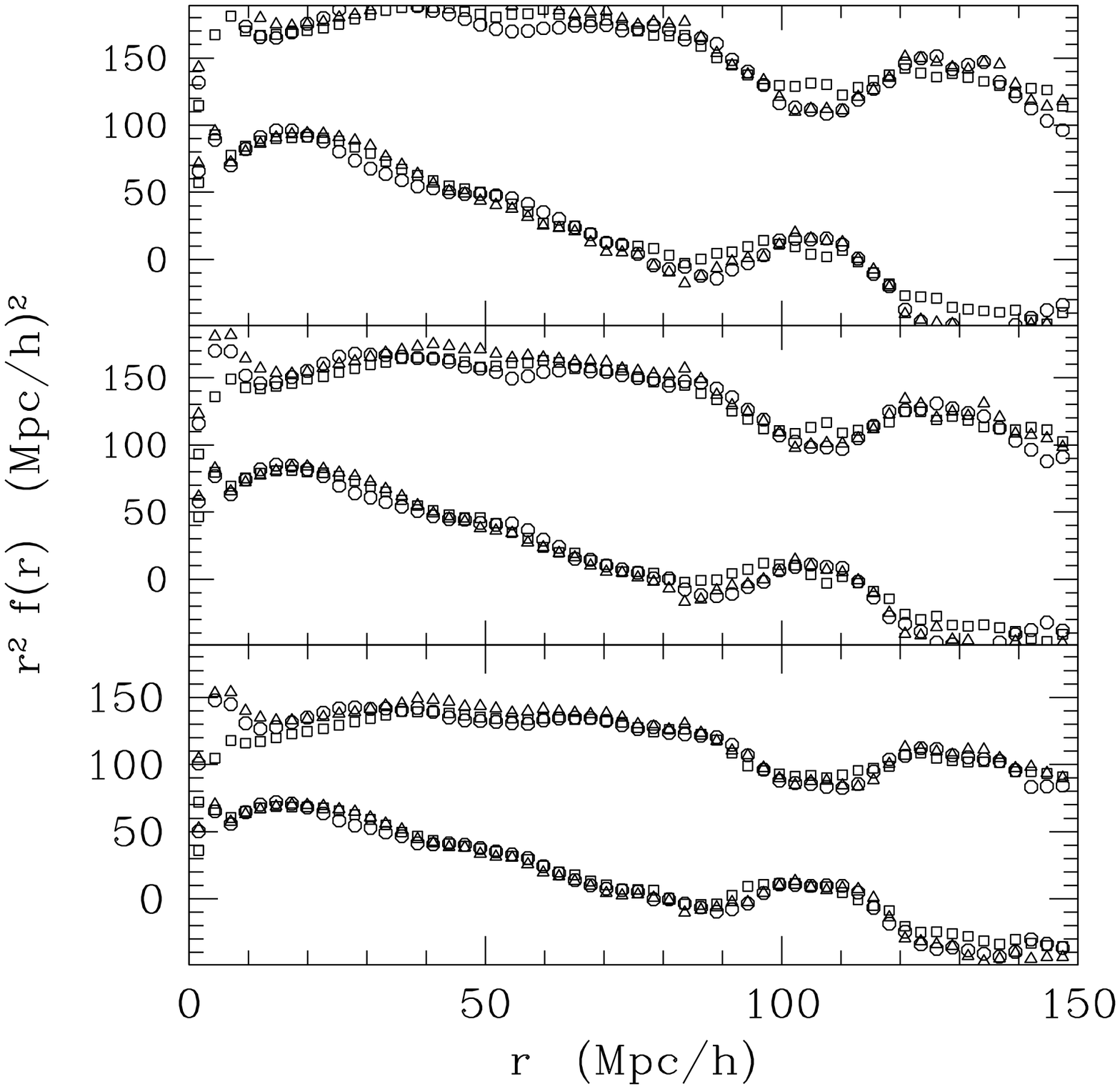}}
\caption{The power spectrum (left) and correlation function (right) for one
of our boxes at 3 different redshifts, $z=0.3$, 0.5 and 1.0, when using
the reconstruction method of Ref.~\protect\cite{ESSS06}.  For the correlation
function we show both $\xi(r)$ (lower curves) and $\Delta\xi(r)$ (upper curves).
For the power spectrum we have plotted the signal compared to the
no-oscillation form of Ref.~\cite{EisHu99}.
The squares are the original signal, the triangles show a reconstruction using
a smoothed field with $R_s=20\,h^{-1}$Mpc and the circles with
$R_s=10\,h^{-1}$Mpc.}
\label{fig:es3}
\end{figure}

The original work of \cite{ESSS06} did not make their measurements on `galaxy'
catalogs made by populating halos with a range of occupation distributions.
We have cross-checked their method using some of our catalogs and found very
consistent results.  We show the results from one of our simulations and
models in Fig.~\ref{fig:es3} in both configuration and Fourier space.  We
present the results here without redshift space distortions to better show
the degree to which reconstruction gains signal.
In agreement with \cite{ESSS06} we find that the real space correlation
function at low redshift is considerably sharpened by the reconstruction
method.  The size of the effect is reduced as we go to higher $z$, and
by $z=1$ the gains are significantly less pronounced.

Since the reconstruction procedure is inherently non-linear, we also tested
whether it induces correlations between otherwise uncorrelated modes.  To
do this we used the 100 non-linearly processed Gaussian density fields
described in \S\ref{sec:sim}.
For each we computed the power spectrum before and after reconstruction and
hence the covariance matrix.  On the scales of relevance for the acoustic
oscillations the procedure does not seem to introduce significant correlations.
Further investigations of reconstruction will be deferred to a future
publication.

\section{Conclusions} \label{sec:conclusions}

The coupling of baryons and photons by Thomson scattering in the early
universe leads to a rich structure in the power spectra of the CMB
photons and the matter.  The study of the former has revolutionized
cosmology and allowed precise measurement of a host of important
cosmological parameters.  The study of the latter is still in its
infancy, but holds the potential to constrain the nature of the dark
energy believed to be causing the accelerated expansion of the universe.

Future large redshift surveys offer the opportunity to measure a
characteristic scale in the universe: the sound horizon at the time
of photon-baryon decoupling.  This standard ruler, which we can calibrate
{}from observations of the CMB, may allow us to tightly constrain the
evolution of the scale factor and determine the nature of dark energy.
To ensure the success of these efforts we need to improve our understanding
of the theoretical underpinnings of the method and generate simulated
universes which can be used to refine and test our observational strategies.

In this paper we have made a first attempt to go all the way from (mock)
observations to constraints on the sound horizon for a number of galaxy
catalogs which display non-linear, scale dependent and stochastic bias.
We have performed fits in configuration and Fourier space for a number of
models which have been proposed in the literature.  We investigate the
shape of the likelihood function, parameter degeneracies and the range of
validity of the fits.  We find that the forms of
Eqs.~(\ref{eqn:lrg}-\ref{eqn:esw}) fare quite well and lead to unbiased
estimates of the sound horizon in both real and redshift space.
In agreement with earlier work \cite{Fisher}, we find that a survey of
several Gpc${}^3$ would constrain the sound horizon at $z\sim 1$ to
about 1\%.

We would like to thank Istvan Szapudi for helpful conversations on
correlation function estimators and edge correction, Ravi Sheth for
numerous enlightening conversations on a number of issues and Joanne
Cohn and Eric Linder for conversations and comments on the manuscript.
MJW also thanks the staff of the Aspen Center for Physics for their
hospitality while part of this work was completed.
The analysis reported here was done on computers at NERSC and LANL.
This work was supported in part by NASA and the NSF.

\end{document}